\documentclass[aps,pre,twocolumn,groupedaddress,10pt]{revtex4-1}

\usepackage[utf8]{inputenc}
\usepackage[T1]{fontenc}
\usepackage{amsmath}
\usepackage{amssymb}
\usepackage{graphicx}
\usepackage{color}
\usepackage{verbatim}
\usepackage{float}

\newcommand{\beq}{\begin{equation}}
\newcommand{\eeq}{\end{equation}}

\usepackage{libertine}
\usepackage[libertine]{newtxmath}

\begin{document}


\title{Token-driven totally asymmetric simple exclusion processes}


\author{Bor Kav\v ci\v c}
\author{Ga\v sper Tka\v cik}
\email[]{gasper.tkacik@ist.ac.at}
\affiliation{Institute of Science and Technology Austria, Am Campus 1, AT-3400 Klosterneuburg, Austria}


\date{\today}

\begin{abstract}
We consider a family of totally asymmetric simple exclusion processes (TASEPs), consisting of particles on a lattice that require binding by a ``token'' in various physical configurations to advance over the lattice. Using a combination of theory and simulations, we address the following questions: (i)~How token binding kinetics affects the current-density relation on the lattice; (ii)~How this current-density relation depends on the scarcity of tokens; (iii)~How tokens propagate the effects of the locally-imposed disorder (such as a slow site) over the entire lattice; (iv)~How a shared pool of tokens couples concurrent TASEPs running on multiple lattices; (v)~How our results translate to TASEPs with open boundaries that exchange particles with the reservoir. Since real particle motion (including in biological systems that inspired the standard TASEP model, {\sl e.g.,} protein synthesis or movement of molecular motors) is often catalyzed, regulated, actuated, or otherwise mediated, the token-driven TASEP dynamics analyzed in this paper should allow for a better understanding of real systems and enable a closer match between TASEP theory and experimental observations.
\end{abstract}

\maketitle


%

\section{Introduction}
Out-of-equilibrium systems involve sustained currents of particles or energy. Totally asymmetric simple exclusion process (TASEP) is an archetypal example of non-equilibrium physics used to describe processes ranging from processive molecular systems to vehicular traffic. Motivated initially as a model of biopolymer synthesis~\cite{macdonald68,macdonald69}, TASEP considers particles that move stochastically on a 1D lattice and obstruct each other's movement by hard-core repulsion (exclusion). Total asymmetry of particle movement is an extreme case in which particles move exclusively in one direction. 

In the basic TASEP formulation, particles hop forward on their own. Yet, real phenomena that motivate the TASEP model often require an external ``movement-enabling'' token  to be present on the particle before the particle can attempt a hop. For example, during protein synthesis ribosomes move forward only when bound by specific proteins called translocation factors~\cite{rodnina18}. Myosin molecular motors similarly require binding of ATP, which provides the necessary energy for motion~\cite{alberts02}. 

In these example cases and more generally, tokens can have distinct association and dissociation characteristics as they interact with the hopping particles. Sometimes, token-particle interactions can be fully accounted for by simple low-order chemical kinetics; alternatively, such kinetics can be more complicated, involving refractory periods or allostery. The effects of these complications, as well as other types of disorder on the lattice ({\sl e.g.,} a ``slow site'' on the lattice which impedes the particle current~\cite{janowsky92}), can be potentiated when the reservoir of tokens is finite, and thus the binding of a token to a particle depletes the pool of tokens available for other hopping particles. In line with these considerations, the first part of our paper is dedicated to elucidating the effects of token-particle interactions and token supply on the overall TASEP dynamics. Previous studies~\cite{brackley10,brackley12,ciandrini10,basu07} considered the effect of limited resources through abstracting the token-dependent hopping as a varying (hopping) rate as a consequence of limited abundance and ``charging'' of tokens. Alternative way of considering the effect of limited-resources is by modeling particle's hopping as a two- or multi-cycle process \cite{klumpp08c}, thus abstracting the binding and consequential hopping actuation. Here, we explicitly focus on the effects of token binding and its interplay with particle hopping. 

In the second part of the paper we focus on the fact that experimentally observable TASEPs are usually open systems. Particles or tokens are therefore exchanged with the environment and possibly shared between TASEP processes running concurrently on multiple lattices. To illustrate this point, we first consider two TASEPs on a ring with a shared finite pool of tokens; we show that coupling through the shared token pool can propagate fluctuations from one TASEP to another. We conclude our treatment by studying truly open TASEP processes that exchange particles with a reservoir via finite entry and exit rates. 

This study uses token-driven TASEP as an example of how finite resources required for particle movement reshape the expected dynamics of stochastically progressing particles with steric exclusion. Our goal is  to provide a theoretical compendium of possible token-driven TASEP model extensions and refinements, and analyze their impact in terms of macroscopic effects. We provide further motivating examples for various extensions in the corresponding sections, and return to the interpretation of  results in the context of observable systems in the Discussion section.

\section{Token-driven TASEP on a ring}
\begin{figure}
\centering
   \includegraphics[width=8.6cm]{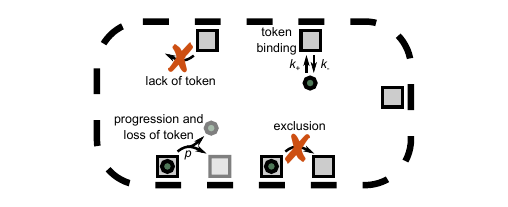}
   \caption{{\bf TASEP on a ring.} Particles (squares) hop from site to site (solid black lines) with rate $p$ if bound by a token (green circles) and if there is an empty site in front. Exclusion prevents particle progression even if the particles are bound by a token. Particles acquire or lose tokens with rates $k_+$ and $k_-$, respectively.
    }
   \label{fig:system}
\end{figure}

We begin the analysis by studying the case of token-driven TASEP with periodic boundary conditions~(Fig.\ref{fig:system}). As the system is closed, the number of particles on the lattice does not change. The average density of particles $\rho$ therefore becomes a control parameter that governs the current-density relation which we derive in what follows. For simplicity, we assume that particles occupy a single site on a lattice rather than extend over several sites. The lattice has $L$ sites in total, and in the assumed ring geometry, the $L$-th site connects to the first site.

In the standard TASEP, in which particles hop forward with the rate $p$, the current-density relation reads
\beq
J=\rho(1-\rho)p.\label{eq:cdr_basic}
\eeq
This relation can be derived by noting that the average current from site $i$ to $i+1$ is given as 
\beq
J_i=\langle p\sigma_i(1-\sigma_{i+1})\rangle.
\label{eq:current0}
\eeq
Here, $\sigma_i$ is 0 (or 1) for particle absent (or present) on the $i$-th site. Brackets $\langle\dots\rangle$ denote a time average; Eq.~(\ref{eq:current0}) then relates the current to the presence of a particle on $i$-th site and presence of the hole (absences of the particle) on the site next to it. Here, density $\rho_i=\langle \sigma_i\rangle$. If we assume translational invariance, {\sl i.e.,} $\rho_i=\rho~\forall i$, we can write in the mean-field  $\langle \sigma_i(1-\sigma_{i+1})p\rangle\approx\langle \sigma_i\rangle\langle(1-\sigma_{i+1})\rangle p=\rho(1-\rho) p$. In the last step, we assumed that correlations are absent. This simplification leads to the standard result, which describes the time-averaged current in steady state [Eq.~(\ref{eq:cdr_basic})]. However, in the thermodynamic limit ($L\to \infty$), the expression Eq.~(\ref{eq:current0}) is exact~\cite{schuetz01}. On notation: We derive the expressions with explicitly spelled-out rates. However, in simulations, the time is measured in the units of $1/p$, and all rates in units of $p$. Therefore, if not spelled out explicitly, the unit is implied and $p=1$. Explicit use of all parameters should facilitate comparison to the experimental observations.

Even as a relatively simple driven system, TASEP remains challenging for analytical treatment. Decades after its introduction~\cite{macdonald68}, numerous variants of TASEP were described, as were its quantitative properties, including the complete Bethe ansatz-based analytical solution for the master equation for a TASEP on a ring~\cite{priezzhev03}.

Minor breaking of the model's symmetries ({\sl e.g.,}~by including site-specific rates) or addition of new dynamical features ({\sl e.g.,}~tokens) makes obtaining exact results challenging. Thus, to obtain a quantitative understanding of TASEP generalizations, we resort to approximate mathematical treatments in which some aspects of the process are neglected (for example, correlations in the mean-field approach).

\subsection{Infinite pool of tokens}
To analyze how tokens might affect the movement of particles, we loosely follow Ref.~\cite{waclaw19}, which considered TASEP with dynamical disorder. In the standard TASEP, the lattice is occupied by a single kind of particle, but with tokens, this reasoning changes: the lattice now bears particles of two kinds, those with and those without a bound token~(Fig.\ref{fig:system}). We thus designate the particles with two separate symbols, {\sl i.e.,}~$\sigma_i$ and $\nu_i$ for token-free and token-bound particles, respectively. Here, $\nu_i=1$ if the site $i$ is occupied by a particle bound by a token and $0$ otherwise (and analogously for $\sigma_i$). These subpopulations of particles are constrained by $\rho_i=\sigma_i+\nu_i$ (for which $\rho=\langle \rho_i\rangle$), TASEP hopping dynamics, and the kinetics of binding and unbinding transitions, $\sigma_i\leftrightarrow \nu_i$. Since only a token-bound particle can hop forward, the equation for the current is
\beq
J_i=\langle p\nu_i(1-\rho_{i+1})\rangle.
\label{eq:current}
\eeq

Let the tokens bind to and unbind from the particles with rates $k_+$ and $k_-$, respectively. When bound by a token, the particle moves forward with the rate $p$ if the next site is free. We additionally assume that the particle loses the token immediately after the hop: $\nu_i=1\to\nu_{i+1}=0$ while $\sigma_i=0\to\sigma_{i+1}=1$. In this section, we also assume an infinite pool of tokens, which ensures that binding and unbinding rates remain fixed. 

Within this framework we can derive the mean-field expressions. We first estimate the $\langle\nu_i\rangle$ from dynamical equation for $\nu_i$:
\beq
\dot{\nu_i}=\sigma_ik_+ -\nu_i \left[	k_- +p(1-\rho_{i+1})\right].
\label{eq:tokenKinetics}
\eeq
The first term  accounts for token binding to unbound particles. The second term describes unbinding, either due to (i)~particle hopping forward with rate $\nu_i p(1-\rho_{i+1})$, or (ii)~spontaneous unbinding with rate $\nu_i k_-$. Since $\sigma_i=\rho_i-\nu_i$, we solve the kinetic equation for the steady state. This leads to an approximate expression 
\beq
\langle \nu_i\rangle\approx\frac{k_+\rho}{k_- + p(1-\rho)+k_+},
\label{eq:averageNu}
\eeq
where we replaced $\rho_{i}$ with the average density $\rho$. $\langle \nu_i\rangle$ is density-dependent and approaches the Langmuir form $k_+/(k_++k_-)$ only as $\rho\to 1$.

We further determine the steady-state value of $\langle \sigma_i\rangle$. The dynamical equation reads:
\beq
\dot{\sigma_i}=k_-\nu_i +\nu_{i-1}(1-\rho_i)p -k_+\sigma_i,
\eeq
where the terms correspond to token unbinding, particle hopping from site $i-1$, and token binding, respectively. Noting that $\nu_i=\rho_i-\sigma_i$ and assuming $\langle\rho_i\rangle=\rho~\forall i$, $\sigma_i=\sigma~\forall i$, we arrive to:
\beq
\langle\sigma_i\rangle\approx\frac{\rho\left[k_- + p(1-\rho)	\right]}{k_++k_- +p(1-\rho)}.
\label{eq:eq:averageSigma}
\eeq
By summing Eqs.~(\ref{eq:averageNu}) and (\ref{eq:eq:averageSigma}), we can verify that the mean-field dynamics obeys the conservation of particles, {\sl i.e.,}~$\langle\sigma\rangle+\langle\nu\rangle=\rho$.  

Mean-field expression for the current [Eq.~(\ref{eq:current})] can be estimated by omitting the effects of correlations:
\beq
J_0\approx p\langle \nu_i\rangle\langle (1-\rho)\rangle=p\langle \nu\rangle(1-\rho),
\label{eq:j0-periodic}
\eeq
where we replaced $\langle \nu_i\rangle$ with $\langle \nu\rangle$ due to translational invariance. With this expression in hand, we solve for the position of maximal current by setting $\partial_\rho J_0=0$:
\beq
\rho_{\rm max}=\frac{	k_-+k_++p-\sqrt{\left(	k_-+k_+\right)\left(	k_-+k_++p\right)}	}{p},
\label{eq:maxCurrent_infinite}
\eeq
which yields $\rho_{\rm max}=1/2$ when $k_+\rightarrow \infty$, as expected for the standard TASEP with $J=\rho(1-\rho)p$. 

\begin{figure}[t!]
\centering
   \includegraphics[width=8.6cm]{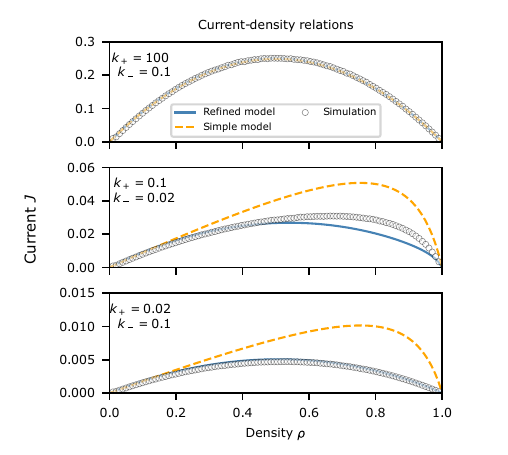}
   \caption{{\bf Current-density relations for token-driven TASEP on a ring in an infinite token pool.} Dependence of current on particle density changes with the kinetic rates $k_+$ and $k_-$. Rapid binding ($k_+\gg p, k_-$, top) results in the standard relation [Eq.~(\ref{eq:cdr_basic})] between $\rho$ and $J$ that both the simple [Eq.~(\ref{eq:j0-periodic})] and the refined [Eq.~(\ref{eq:j1-periodic_general})] models predict correctly. Reducing $k_+,k_-$ lowers the maximal current and skews the relation towards higher $\rho$ (middle), which is less pronounced if $k_+<k_-$ (bottom). Note: simulation data completely covers the prediction of the refined model in the bottom panel and thus we made plot markers semi-transparent.
    }
   \label{fig:rho-J-infinite}
\end{figure}

Alternatively, we can derive another current estimate by considering the typical time between hops for a randomly chosen particle~\cite{waclaw19}. In this scenario, the ``refined'' approximation for the current is:
\beq
J_1=\bar\tau^{-1}\rho(1-\rho),
\label{eq:j1-periodic_general}
\eeq
where $\bar\tau$ is the typical waiting time for a given particle to move in the absence of exclusion. In Eq.~(\ref{eq:j1-periodic_general}), $\rho$ is the probability of finding any particle while $1-\rho$ weights the effect of exclusion. In general, calculating $\tau$ is equally challenging as is the exact evaluation of the current [Eq.~(\ref{eq:current})]. However, we can approximate $\tau$ by considering a sequence of events that need to transpire for the particle to move, and the probability of each such event.

Estimation of the typical time $\tau$ becomes considerably simpler if we assume that the hopping rate $p$ is high enough that as soon as the particle is bound, it will attempt jumping forward. Combined with a separate assumption that $k_-<k_+$, it becomes unlikely that the particle, once bound, would undergo several binding-unbinding events. 
Under these assumptions the typical waiting time for a given particle reads:
\beq
\bar\tau\approx\frac{\langle\nu\rangle}{p\rho} + \left(1-\frac{\langle\nu\rangle}{\rho}\right)\left(\frac{1}{k_+} +\frac{1}{p}\right).
\eeq
The first term describes the hopping of a particle forward when bound (weighted by the conditional probability of being bound given the particle at the site, $\langle\nu\rangle/\rho$), while the second term describes the events where the particle is first unbound (with conditional probability $1-\langle\nu\rangle/\rho$), upon which it acquires a token and hops forward. The resulting current estimate combines into:
\begin{eqnarray}
J_1&\approx&\frac{\rho k_+p}{\rho(p+k_+)-\langle\nu\rangle p}\rho(1-\rho)\label{eq:j1}= \label{eq:j1-periodic}\\
&=&\frac{\rho k_+p}{\rho(p+k_+)- p{k_+\rho}/\left[k_- + p(1-\rho)+k_+\right]}\rho(1-\rho).  \nonumber
\end{eqnarray}
The prefactor in Eq.~(\ref{eq:j1}) captures the deviation from the standard TASEP relation~[Eq.~(\ref{eq:cdr_basic})].

We compare these theoretical predictions to stochastic TASEP simulations. In simulations, we varied the density $\rho$ and estimated the current~(see Appendix~\ref{sec:simulations}) for different $k_+$ and $k_-$~(Fig.~\ref{fig:rho-J-infinite}). We first confirm that in the limit $k_+\to \infty$, we recover the standard density-current relation [Eq.~(\ref{eq:cdr_basic})], which is correctly predicted by the simple mean-field [Eq.~(\ref{eq:j0-periodic})] and the refined model [Eq.~(\ref{eq:j1-periodic_general})]. However, if rates $k_+,k_-$ are lowered, the kinetics of token binding causes $J(\rho)$ to deviate away from the standard shape. The simple model  correctly predicts the general direction of skewing, but fails to describe the relation quantitatively, while the refined model performs significantly better. We further find that these simulation results are consistent across different lengths of the lattice $L$ (see Appendix~\ref{sec:length}).

Altering the kinetic rates breaks the symmetry in the current-density relation: the maximum of $J(\rho)$ starts to skew towards higher densities ($\rho>0.5$). Intuitively, the increased likelihood of token binding at higher particle density increases the effective progression rate. This reasoning is, however, valid only if the token pool is infinite, {\sl i.e.,}~when the rates of binding remain constant and independent of the number of already bound tokens.

\subsection{Finite pool of tokens}
\begin{figure}
\centering
   \includegraphics[width=8.6cm]{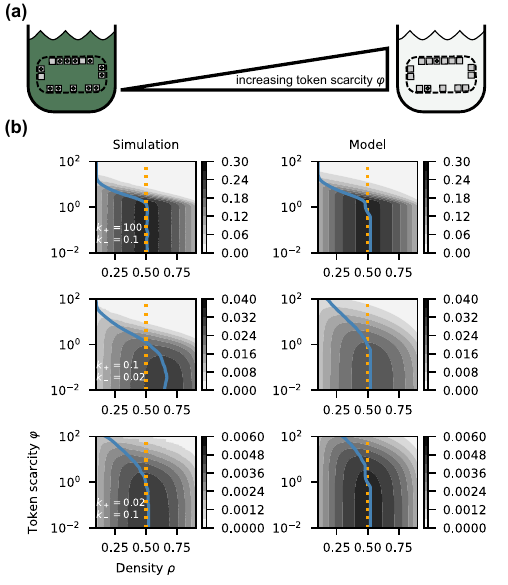}
   \caption{{\bf Token scarcity influences the current-density relation.} (a)~Token scarcity $\varphi$ quantifies total number of tokens relative to the number of lattice sites~[Eq.~(\ref{eq:tokens_finite})]. (b)~Current-density relation surfaces for a finite token pool: simulation (left column) and analytical prediction (right column). Current is depicted in grayscale (bars at right). Shown are three examples with different binding parameters. Blue lines denote the point of maximal current at a given $\varphi$; smoothed by LOESS to mitigate parameter discretization grid. Orange dashed line denotes the position of current maximum for standard TASEP at $\rho=1/2$. Both model and simulation results were evaluated on a $20\times100$ grid. Detailed cross-sections of the surfaces are in Fig.~\ref{fig:periodic_finite_density-rho_examples_plot}.}
   \label{fig:periodic_finite_surfaces}
\end{figure}

We next consider the case where the number of tokens available for binding is finite~(Fig.~\ref{fig:periodic_finite_surfaces}a). We model this situation as follows:
\begin{eqnarray}
\dot{\nu_i}&=&\sigma_i\left(\frac{N_0-\langle\nu\rangle L}{N_0}\right)k_+ -\nu_i\left[k_- +p(1-\rho_{i+1})\right]=\nonumber\\
&=&\sigma_i\left(1-\langle\nu\rangle\varphi\right)k_+ -\nu_i\left[k_- +p(1-\rho_{i+1})\right],\label{eq:tokens_finite}
\end{eqnarray}
where $N_0$ is the total number of tokens. The term $(N_0-\langle\nu \rangle L)/N_0=\left(1-\langle\nu\rangle\varphi\right)$ determines the fraction of free tokens relative to the total pool $N_0$; we call $\varphi=L/N_0$ the ``token scarcity''~(Fig.~\ref{fig:periodic_finite_surfaces}a). When $\varphi=0$, token binding does not deplete the pool of free tokens and we recover the infinite token case~[Eq.~(\ref{eq:tokenKinetics})]. Increasing $\varphi$ towards $1$ brings us to the situation where there is exactly one token per lattice site, while $\varphi\to\infty$ completely depletes the free tokens. In steady state we find:
\begin{eqnarray}
\langle \nu \rangle_\pm&\approx&
\frac{2 k_+\rho}{\Psi\pm\sqrt{\Psi^2-4k_+^2\rho\varphi}}\label{eq:nuFinite}\\
\text{with } \Psi&=&k_+\left(\rho\varphi+1\right) + k_- +\left(1-\rho\right)p,\nonumber
\end{eqnarray}
where the solution with the positive sign yields a physically realistic result. We recover Eq.~(\ref{eq:averageNu}) as $\varphi\to 0$. 

For unbound particles,
\beq
\dot{\sigma_i}=k_-\nu_i +\nu_{i-1}(1-\rho_i)p -k_+\sigma_i\left(1-\varphi \langle\nu\rangle\right)
\eeq
yields, in stady state,
\beq
\langle\sigma\rangle_\pm\approx\frac{2\gamma}{\Delta\pm\sqrt{\Delta^2-4\varphi k_+\gamma}},
\eeq
where $\Delta=k_-+p(1-\rho)+k_+(1-\varphi\rho)$ and $\gamma=-\rho\left[	k_-+p(1-\rho)	\right]$. Only the negative-sign solution is physically realistic. From Eq.~(\ref{eq:nuFinite}) we recognize that $\Psi^2-4k_+^2\rho\varphi=\Delta^2-4\varphi k_+\gamma$, which helps us confirm that particles are conserved, $\langle \sigma\rangle+\langle\nu\rangle=\rho$.

Mean-field  approximation for the current [Eq.~(\ref{eq:j0-periodic})] remains the same, but the refined model [Eq.(~\ref{eq:j1-periodic_general})] requires a modification. Since the binding rates depend on the abundance of free tokens, the typical waiting time must be altered:
\begin{eqnarray}
\bar\tau&\approx&\frac{\langle\nu\rangle}{p\rho} + \left(1-\frac{\langle\nu\rangle}{\rho}\right)\left[\frac{1}{ k_+\left(1-\varphi \langle\nu\rangle\right)} +\frac{1}{p}\right]=\nonumber\\
&=&\frac{1}{p}+\frac{\rho-\langle \nu \rangle}{ \rho k_+\left(1-\varphi \langle\nu\rangle\right)}.\label{eq:tauFinite}
\end{eqnarray}
Plugging this expression for $\bar \tau$ into Eq.~(\ref{eq:j1-periodic_general}) yields an approximate expression for the current which takes into account the limited token pool:
\beq
J_1\approx
\frac{(1-\varphi \langle\nu\rangle)  \rho k_+p}{p(\rho-\langle\nu\rangle) + k_+(1-\varphi \langle\nu\rangle)\rho}
\rho(1-\rho).
\label{eq:J1_finite}
\eeq

This expression allows us to predict the current as a function $J(\rho,\varphi)$, as shown in Fig.~\ref{fig:periodic_finite_surfaces}b (see Appendix~\ref{sec:density-current_relations} for \mbox{current-density} relations in more detail). 
The degree of token scarcity $\varphi$ profoundly affects the current-density relation. As $\varphi$ increases beyond $1$, we note that the maximal current is reached at lower densities. Importantly, this inverts the behavior we have observed for an infinite token pool, where the maximal current occurs at $\rho>0.5$. Token scarcity is less limiting if tokens that bind are likely to bind to particles that are free to hop, thus favoring low particle density. On the other hand, lowering the density too far reduces the overall current as well. This tradeoff causes the current to peak at a location determined by $k_+$, $k_-$, and by the availability of tokens. Our analytical expression correctly predicts the qualitative changes in the maximal $J(\rho)$ as a function of~$\varphi$. We also show that for $k_+=0.1$ and $k_-=0.02$ these results largely do not depend on the lattice length (see Appendix~\ref{sec:length}).

When tokens are abundant, $\varphi\sim0$, both limits $k_+\to0$, $k_+\to \infty$ give a symmetric current-density relation [Fig.~\ref{fig:periodic_finite_surfaces}]. To understand the behavior at low $k_+$, we expand Eq.~(\ref{eq:J1_finite}) into a series around $k_+=0$, which yields $J\approx k_+\rho(1-\rho)$. This equation is the same as for the standard TASEP, when the hopping rate $p$ is replaced by the binding rate $k_+$. Intuitively, when all rates are much higher than $k_+$, any particle bound by a token (a process limited by $k_+$), will hop forward. Thus, we can quickly estimate the maximal current at low binding rates: $J_{\rm max}\approx k_+/4$, which we can verify by inspecting Figs.~\ref{fig:rho-J-infinite} and \ref{fig:periodic_finite_surfaces}. In the remainder of this paper, we focus on binding parameters outside of the two limiting $k_+$ regimes.

\subsection{Token-retention and the refractory scenarios} 
   
\begin{figure*}
\centering
   \includegraphics[width=17.2cm]{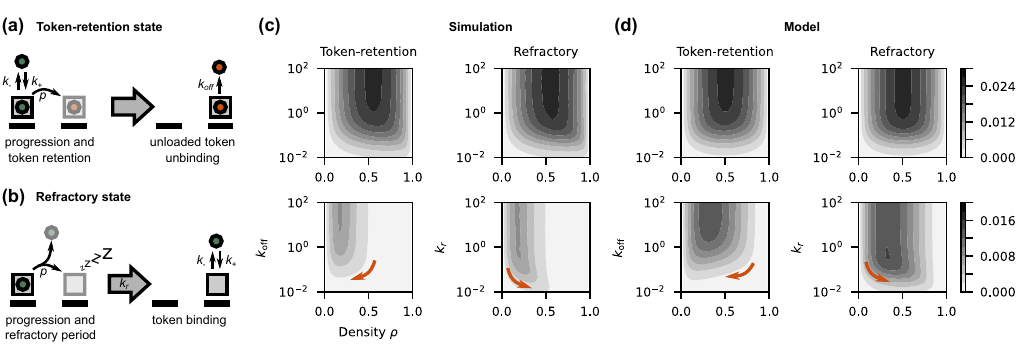}
   \caption{{\bf Token-retention and refractory scenarios.} (a)~Token-retention scenario: Tokens in a loaded state (green circles)  bind reversibly to the particle (squares). If a particle progresses by making a hop, its token unloads in the process (red circles), and later unbinds with a rate $k_{\rm off}$ that can differ from $k_-$. Upon detaching, tokens enter into the token pool and reload. (b)~Refractory scenario: As in~(a), tokens bind to the particles reversibly. Upon progression, particles lose the token and enter into a refractory period of characteristic duration $\tau_{r}=1/k_{r}$. (c)~Impact of token-retention and refractory scenarios on the current-density relation for different token scarcities $\varphi$. Left column: current (gray scale) for a token-retention TASEP for different degrees of token scarcity $\varphi$ (top: abundant tokens, $\varphi\approx0.78$; bottom: scarce tokens, $\varphi\approx14.38$). Right column: same as left, but for refractory TASEP. When tokens are abundant and $\varphi\to 0$, the behavior of the two scenarios is qualitatively the same, differences emerge as $\varphi$ increases beyond 1. Token-retention scenario leads to the depletion of the token pool for low values of $k_{\rm off}$, while the refractory scenario does not. Decreasing $k_r$ transiently increases the current at a given density in the refractory scenario but not for the retention scenario (see arrows). Results obtained for $k_+ = 0.1$ and $k_- = 0.02$. (d)~As in~(c), but for model results. Detailed cross-sections yielding density-current relations are in Fig.~\ref{fig:twostage_supp}.
    }
   \label{fig:twostage}
\end{figure*}

We now focus on the situation in which tokens do not instantly detach from the particle after the particle makes a hop. For example, a ``loaded'' token could bind reversibly to a particle, actuate the movement of the particle (becoming ``unloaded'' or spent in the process) and then remain bound on the particle in an ``unloaded'' state until it unbinds~(Fig.~\ref{fig:twostage}a).
When tokens are ``unloaded'', they can detach with a characteristic rate $k_{\rm off}$ from particles but cannot re-load; re-loading is only possible by returning to the token pool. In this ``token-retention'' scenario, tokens remain bound to the particles after the latter have hopped forward, thereby acting as a sponge that sequesters free tokens.

Alternatively, the particles lose the token immediately after making a hop, upon which they enter into a state where they are forbidden from binding a new token for a typical time $\tau_r=1/k_{r}$ ($k_r$ is the exit rate from refractory state). We call this scenario ``refractory''~(Fig.~\ref{fig:twostage}b), in analogy to similar phenomena seen in neurons or pacemaker cells in the heart; there, cells can only fire a new spike after a sufficient time has passed from the previous spike. Since the tokens return back to the pool immediately after a particle's forward hop, the refractory scenario is expected to impact the TASEP dynamics in a different manner to the token retention scenario.

\subsubsection{Token-retention scenario}
To analyze the effects of token retention on the current-density relation, we consider two populations of tokens: those bound in a loaded state before the hop ($\nu$) and those bound in an unloaded state after the hop ($b$). The higher the abundance of tokens in either state, the fewer free tokens remain in the pool to facilitate the movement of unbound particles. The dynamics of $\nu$ obeys:
\begin{eqnarray}
\dot{\nu}&=&\sigma\left[\frac{N_0-L\left(\langle \nu \rangle +\langle b \rangle\right)}{N_0}\right]k_+-\nu \left[	k_- +p(1-\rho)	\right]=\nonumber\\
&=&\sigma\left[	1-\varphi(\langle \nu \rangle +\langle b \rangle)	\right]k_+ - \nu \left[	k_- +p(1-\rho)	\right].
\label{eq:dotnu-TwoStage}
\end{eqnarray}
We note that when $L\left(\langle \nu \rangle +\langle b \rangle\right)/N_0=\varphi(\langle \nu \rangle +\langle b \rangle)=1$, the binding of tokens to particles ceases. 

Because $\rho=\langle \sigma\rangle+\langle \nu\rangle+\langle b\rangle $, we can express $\langle b\rangle $ as function of $\langle \nu \rangle$ by solving 
\beq
\dot b=\nu p(1-\rho)-k_{\rm off}b,
\eeq
where $k_{\rm off}$ is the rate of token unbinding after a particle hop. The steady-state solution reads:
\beq
\langle b\rangle\approx\frac{p(1-\rho)}{k_{\rm off }}\langle \nu \rangle,
\eeq
which allows us to write $\langle \sigma\rangle=\rho-\langle \nu\rangle \left[	1+p(1-\rho)/k_{\rm off }	\right]$. Plugging this expression into Eq.~(\ref{eq:dotnu-TwoStage}) allows us to derive the steady-state solution for $\nu$:
\begin{eqnarray}
\langle \nu \rangle &\approx& \frac{2\rho k_+}{\Psi_{\rm off}\pm\sqrt{\Psi_{\rm off}^2-4k_+^2\varphi\theta^2\rho}}\label{eq:nu2Stage}\\
\text{with } \Psi_{\rm off}&=&k_+\theta\left(\rho\varphi+1\right) + k_- +\left(1-\rho\right)p,\nonumber
\end{eqnarray}
where $\theta=1+p(1-\rho)/k_{\rm off }$. We verify that in the limit $k_{\rm off}\to\infty$ Eq.~(\ref{eq:nu2Stage}) becomes Eq.~(\ref{eq:nuFinite}), since this limit corresponds to instantaneous token unbinding after the particle hop. For $\varphi\to 0$, we obtain the expression for $\nu$ that resembles Eq.~(\ref{eq:averageNu}), but with a rescaled~$k_+\to k_+\theta$.

In the retention scenario, we can derive a refined approximation for the current, following Eq.~(\ref{eq:j1-periodic_general}). To this end, we need an updated expression for the ``typical waiting time'', $\bar{\tau}$, which now sums over three contributions (particles bound and ready to hop forward, particles unbound, or particles bound by a token in unloaded state)s:
\begin{eqnarray}
\bar{\tau}&\approx&\frac{\langle\nu\rangle}{p\rho} + \frac{\sigma}{\rho}\left[\frac{1}{ k_+\left(1-\varphi \theta\langle\nu\rangle\right)} +\frac{1}{p}\right]+\nonumber\\
&+&\frac{b}{\rho}\left[\frac{1}{k_{\rm off}}+\frac{1}{ k_+\left(1-\varphi \theta\langle\nu\rangle\right)} +\frac{1}{p}\right]. \label{eq:tau2Stage}
\end{eqnarray}
Since $\sigma=\rho-\theta\nu$ and $b=\nu(\theta-1)$, we recover Eq.~(\ref{eq:tauFinite}) as $\theta\to 1$, as expected. 

\subsubsection{Refractory period scenario}
In this scenario, we partition the particles into unbound ($\sigma$), bound ($\nu$), and those in a refractory state ($z$). The overall density equals $\rho=\langle \sigma\rangle+\langle \nu\rangle+\langle z\rangle$. We first focus on particles in the refractory state:
\begin{eqnarray}
\dot z&=&\nu p(1-\rho)-k_rz,\label{eq:diff_refrZ}\\
\langle z \rangle &\approx&\frac{p(1-\rho)}{k_r}\langle \nu \rangle\label{eq:refrZ}.
\end{eqnarray}
In Eq.~(\ref{eq:diff_refrZ}), the first term corresponds to particles entering into the refractory state by hopping forward, while the negative term describes the particles leaving the refractory state with a characteristic rate $k_r$. As $k_r\to \infty$, $\langle z \rangle \to 0$, as expected. 

Using Eq.~(\ref{eq:refrZ}) we  rewrite the equation for the overall density as $\rho=\langle \sigma\rangle+\langle \nu\rangle \left[	1+p(1-\rho)/k_r	\right]$, which we use to find the steady state for the differential equation for $\nu$:
\beq
\dot \nu=\sigma(1-\varphi \nu)k_+-\nu\left[ k_-+p(1-\rho)		\right].
\eeq
We find:
\begin{eqnarray}
\langle \nu \rangle &\approx& \frac{2\rho k_+}{\Psi_r\pm\sqrt{\Psi_r^2-4k_+^2\varphi\Omega\rho}}\label{eq:nuRef}\\
\text{with } \Psi_r&=&k_+\left(\rho\varphi+\Omega\right) + k_- +\left(1-\rho\right)p,\nonumber
\end{eqnarray}
where $\Omega=1+p(1-\rho)/k_r$. Because $\lim_{k_r\to \infty}\Omega=1$, Eq.~(\ref{eq:nuRef}) transforms into Eq.~(\ref{eq:nuFinite}) which is valid for the finite token pool with the instant release of tokens upon particle progression. In the limit of infinite tokens, $\varphi\to 0$, the  expression for $\nu$ is similar to Eq.~(\ref{eq:averageNu}) in which~$k_+\to k_+\Omega$. We finally find the ``typical waiting time'' which we can use in the refined estimate for the current [Eq.~(\ref{eq:j1-periodic_general})]: 
\begin{eqnarray}
\bar{\tau}&\approx&\frac{\langle\nu\rangle}{p\rho} + \frac{\sigma}{\rho}\left[\frac{1}{ k_+\left(1-\varphi \langle\nu\rangle\right)} +\frac{1}{p}\right]+\nonumber\\
&+&\frac{z}{\rho}\left[\frac{1}{k_r}+\frac{1}{ k_+\left(1-\varphi \langle\nu\rangle\right)} +\frac{1}{p}\right].\label{eq:tauRef}
\end{eqnarray}
This estimate for $\bar \tau$ is only slightly different from Eq.~(\ref{eq:tau2Stage}); note the absence of the weighing factor $\theta$ in the binding terms.

\subsubsection{Simulation results and analytical approximations}
We compare the analytical approximations for the current based on $\bar\tau$ for both scenarios [Eqs.~(\ref{eq:tau2Stage},\ref{eq:tauRef})] to the stochastic simulations in Fig.~\ref{fig:twostage}c,d (see Appendix~\ref{sec:detailed_open} for density-current relations). For both scenarios we observe a good qualitative match. Firstly, the models correctly predict that there is no difference between both scenarios when $\varphi\sim0$: the particles are unable to bind new tokens either due to being bound by an unloaded token or due to being in refractory state---neither state has an effect on the abundance of tokens when those are present in infinite number. Secondly, the model correctly predicts qualitative difference for $\varphi\gg1$: maximal current for a given $k_{\rm off}$/$k_r$ moves towards lower or higher densities for token-retention or refractory scenarios, respectively. Thus, for scarce tokens, higher current is achieved at higher densities if particles exhibit a refractory period.

\subsection{Effects of a slow-site on token-driven TASEP\label{sec:slow}}
We consider a TASEP on a ring containing a single slow site where the particle hopping rate is locally reduced by a barrier factor $r\leq1$, {\sl i.e.,}~$p\to pr$. If $r=0$, particles cease to progress through the site, while for $r=1$ there is no slow site. Even though the disorder in the lattice is local, the slow site alters the dynamics of the particle progression globally and can thereby substantially affect the average current.
\begin{figure}[b!]
\centering
   \includegraphics[width=8.6cm]{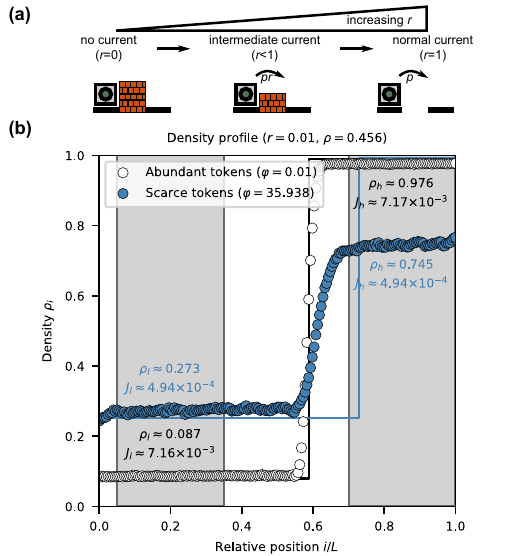}
   \caption{{\bf Density profiles for a single slow site in TASEP on a ring.} (a)~Schematic of a slow site. When the slow site is present, the hopping rate across the slow site decreases by a barrier factor $r \in [0,1]$. This can be interpreted as a fixed ``obstacle'' that particles need to pass over (depicted by a wall). (b)~Examples of density profiles for two different token scarcities (see legend), calculated for $k_+=0.1$ and $k_-=0.02$. The slow site in this example is located at relative position $i/L=1$ and has a barrier factor of $r=0.1$. When tokens are abundant ($\varphi\ll1$), the density profiles consists of two distinct spatial segments on a ring separated by a sharp shock front. When tokens are scarce ($\varphi\gg1$), the densities in the two segments, $\rho_l$ and $\rho_h$, become more similar,  the low density segment expands, and the shock front becomes less sharp. Tiny transition effects in the form of small kinks form before and after the slow site. Shading denotes the segments where we compute the average (low and high) densities and currents; while densities are different, the average bulk currents within both segments must (and do) remain the same. Solid lines are predicted density profiles. 
    }
   \label{fig:periodic_slowsite}
\end{figure}

In large systems of self-actuated particles, the density profile separates into two distinct spatial segments: the segment of high density~($\rho_h$) before the slow site and the segment of low density~($\rho_l$) after the slow site~\cite{janowsky92}. These segments are separated by a shock front. How does this picture, derived within the standard TASEP framework, change for the token-driven TASEP?

We simulate token-driven TASEP with a barrier between the first and the last site of the lattice, {\sl i.e.,}~the hopping rate of the token-bound particle from $i=L$ to $i=1$ is $pr$. Simulations show that the density profile indeed breaks into two segments with different densities separated by a shock front. We define $0\leq\xi\leq1$ as a relative size of low-density segment; correspondingly, $(1-\xi)$ is the size of the high density segment. The value for $\xi$ must fulfill $\rho=\rho_h(1-\xi)+\rho_l\xi$.

Within the mean-field approximation, we assume that the average currents in the bulk of either segment and over the slow site are equal; current further has to satisfy Eq.~(\ref{eq:j0-periodic}). These assumptions yield, respectively:
\begin{eqnarray}
\langle\nu\rangle_h(1-\rho_h)=\langle\nu\rangle_l(1-\rho_l),\label{eq:high-low}\\
r\langle\nu\rangle_h(1-\rho_l)=\langle\nu\rangle_l(1-\rho_l),\label{eq:slowSite}
\end{eqnarray}
where the subscripts denote the segments. Combining the two equations yields $\rho_l=1-(1-\rho_h)/r$, which can be rewritten in terms of $\xi$ as $\rho_h=({\rho-\rho_l\xi})/({1-\xi})$, leading to:
\begin{eqnarray}
\rho_l&=&\frac{(1-\xi)(r-1)+\rho}{r(1-\xi)+\xi}\label{eq:rhoL},\\
\rho_h&=&\frac{r\rho-\xi(r-1)}{r(1-\xi)+\xi}.\label{eq:rhoH}
\end{eqnarray}
By considering $r\to 1$, we obtain $\rho_h=\rho_l=\rho$ and $\xi$ remains arbitrary. In contrast, $r\to 0$ results in $\rho_h=1$, from which we see that $\xi=1-\rho$: particles condense onto a segment of length $L\rho$ sites at the maximal attainable density of 1. 

We next look at the intermediate cases that interpolate between the intuitive limits. Since forward hopping depends on the density within a segment, the kinetic equations will be different for either segment:
\begin{eqnarray}
\dot{\nu}_h&=&k_+\sigma_h\left\{1-\varphi\left[\xi \nu_l+(1-\xi)\nu_h	\right]	\right\}-\nonumber\\
&-&\nu_h\left[k_-+p(1-\rho_h)	\right],\label{eq:nuH}\\
\dot{\nu}_l&=&k_+\sigma_l\left\{1-\varphi\left[\xi \nu_l+(1-\xi)\nu_h	\right]	\right\}-\nonumber\\
&-&\nu_l\left[k_-+p(1-\rho_l)	\right],\label{eq:nuL}
\end{eqnarray}
where the effects at the segment interface have been neglected. In steady state, Eqs.~(\ref{eq:nuH}) and (\ref{eq:nuL}) yield
\begin{eqnarray}
\langle \nu \rangle_{j,\pm}&\approx&
\frac{2 k_+\rho_j}{\Psi^\prime_j\pm\sqrt{\Psi^{\prime2}_j-4k_+^2\rho_j\varphi^\prime_j}}\label{eq:nuHL}\\
\text{with } \Psi_j^\prime&=&k_+\left(\rho_j\varphi^\prime_j+1\right) + k_- +\left(1-\rho_j\right)p,\nonumber
\end{eqnarray}
which is analogous to Eq.~(\ref{eq:nuFinite}). Differences between the low- and high-density segment arise from $\rho_j$ [Eqs.~(\ref{eq:rhoL}-\ref{eq:rhoH})] and~$\varphi^\prime_j$:
\begin{eqnarray}
\varphi^\prime_h=\varphi\left[1-\xi\left(1-r\right)		\right]\label{eq:phiH},\\
\varphi^\prime_l=\varphi\frac{1-\xi\left(1-r\right)		}{r}\label{eq:phiL},
\end{eqnarray}
where, as expected, $\varphi^\prime_h=r\varphi^\prime_l$. By combining Eqs.~(\ref{eq:rhoL}-\ref{eq:rhoH}), Eqs.~(\ref{eq:phiH}-\ref{eq:phiL}) and Eq.~(\ref{eq:nuHL}), we can obtain the equation for the shock front position $\xi$.

\begin{figure}
\centering
   \includegraphics[width=8.6cm]{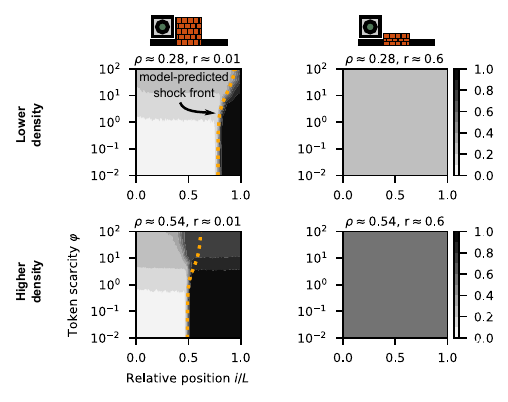}
   \caption{{\bf Emergence and position of the shock front depend on token scarcity, density, and slow site barrier.} Shown are density profiles (grayscale) simulated for  $k_+=0.1$ and $k_-=0.02$. Left (right) column shows strong (weak) slow-site barrier ($r=0.01$ and $r=0.6$, respectively). Top (bottom) row shows low (high) density ($\rho=0.28$ and $\rho=0.54$, respectively). 
   At left ($r=0.01$), the density profile shows how a sharp separation  between segments of different density becomes smoother and ultimately transitions to a uniform  profile, as tokens get increasingly scarce ($\varphi\gg1$). Orange dashed line denotes the numerical solution for shock front position~$\xi=i/L$. At right ($r\sim1$) there is no clear separation between the segments and the analytical approximation returns no physical solution. 
   }
   \label{fig:periodic_slowsite_densities}
\end{figure}

The shock-front equation is not suitable for analytical treatment, but we can solve it using numerical methods~(density profiles in Fig.~\ref{fig:periodic_slowsite}b, shock-front prediction in Fig.~\ref{fig:periodic_slowsite_densities}). The solutions agree qualitatively with the simulations, especially for $\rho<1/2$. Specifically, our approximation correctly predicts the shift of the shock front towards the slow site as token scarcity $\varphi$ increases as well as the increase in the density of the low-density segment (Fig.~\ref{fig:periodic_slowsite}b). The model fails in predicting the exact value of the density in the high-density regime for severe token scarcity~(Fig.~\ref{fig:periodic_slowsite}b). We note that for $r\to1$ the separation between the segments disappears~(Fig.~\ref{fig:periodic_slowsite_densities}). For the standard TASEP, the shock front exists if $|\rho-1/2|<(1-r)/2(1+r)$ as per Ref.~\cite{janowsky92}. This relation does not correctly describe the disappearance of the shock front in our case (see, for instance, Fig.~\ref{fig:periodic_slowsite_densities} where $\rho\approx0.54$ and $r=0.6$). The difference arises because the token-dependent current is lower compared to the standard TASEP, which makes the hopping barrier at the slow site less perturbing. Our approximation cannot account for the finite sharpness of the transition between low- and high-density segments; this would require an explicit account of correlations, which is beyond the mean-field regime.

This example demonstrates that the finite token pool acts as an effective coupling medium through which particles across the whole lattice feel each other. The kinetics of token-particle (un)binding interacts differentially with the low- and high-density segments, so as to shift the balance which determines the shock front position, in deviation from the standard TASEP.

\subsection{Competing TASEPs\label{sec:competing}}
What happens when two TASEP processes share a common pool of tokens? In the slow-site case, the shared pool of tokens coupled the two segments of different density on the same lattice; in this case, the shared pool will couple separate TASEP processes of different density. The increased fraction of token-bound particles on one TASEPs is then expected to limit the availability of free tokens and in turn reduce the binding of the tokens on the other TASEPs.

\begin{figure}
\centering
   \includegraphics[width=8.6cm]{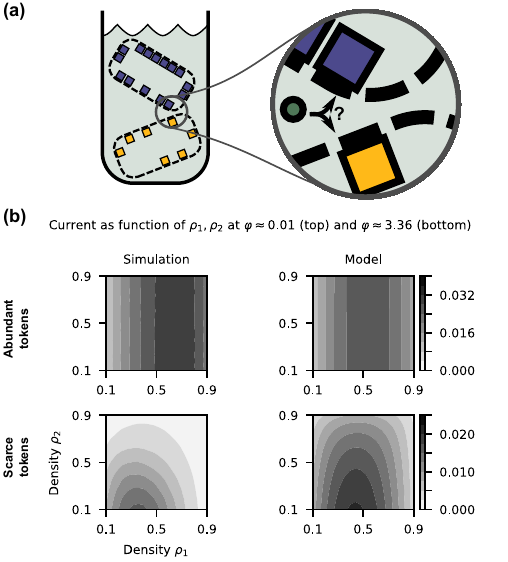}
   \caption{{\bf Two competing TASEPs coupled by a finite shared token pool.} (a)~Schematic showing the competition of particles on two TASEPs for  token binding, which depends on the density of unbound particles on either lattice. (b)~Current of the first TASEP, $J_1$ (grayscale), as a function of particle densities ($\rho_1$, $\rho_2$) on both TASEPs. Simulations (left) and the model (right) compare the results when tokens are abundant (top, $\varphi=0.01$) and when they are scarce (bottom, $\varphi=3.36$). Kinetic rates are fixed at $k_+=0.1$ and $k_-=0.02$.
   }
   \label{fig:periodic_competition}
\end{figure}

We simulate a pair of TASEPs of the same length $L$ with periodic boundary conditions and a shared token pool~(Fig.~\ref{fig:periodic_competition}a). We vary the density of particles on both lattices ($\rho_1$ and $\rho_2$) at different levels of token scarcity $\varphi$ to examine how the current-density relation in one TASEP may be affected by the density of the other TASEP~(Fig.~\ref{fig:periodic_competition}b, left). 

When tokens are abundant ($\varphi\ll1$), the current-density relation of the first TASEP, $J_1(\rho_1)$, remains independent of the density of particles on the second TASEP, $\rho_2$, as expected. When tokens are scarce ($\varphi>1$), in contrast, the current-density relation of the first TASEP is affected profoundly by $\rho_2$. Increasing density $\rho_2$ causes the maximal current of the first TASEP to occur at lower densities $\rho_1$ as the pool of free tokens becomes limiting. 

To obtain analytical insight, we examine the dynamics of $M$ competing TASEPs whose kinetic terms depend on the availability of free tokens in the shared pool:
\begin{eqnarray}
\dot{\nu}_{i,j}&=&\sigma_{i,j}\left(	\frac{N_0 - \sum_{j^\prime}^{M}L_{j^\prime}\langle \nu\rangle_{j^\prime}}{N_0}	\right) k_{+,j}-\nonumber\\
&-&\nu_{i,j}\left[k_{-,j}+p_j\left(	1-\rho_{i+1,j}	\right)	\right]=\nonumber\\
&=&\sigma_{i,j}\left(	1- \varphi\sum_{j^\prime}^{M}\ell_{j^\prime}\langle \nu\rangle_{j^\prime}	\right) k_{+,j}-\nonumber\\
&-&\nu_{i,j}\left[k_{-,j}+p_j\left(	1-\rho_{i+1,j}	\right)	\right],
\label{eq:generalCompet}
\end{eqnarray}
where $j$ indexes different TASEP processes, $i$ indexes the sites of a given TASEP, $\varphi=L/N_0$, and $\ell_{j^\prime}=L_{j^\prime}/L$ is relative length. This general formulation allows for different binding kinetics across TASEPs (as noted by indices of $k_{+,j},k_{-,j}$) as well as different hopping rates $p_j$. For tractability and comparison to the simulations, we now simplify to~$M=2$ and $L_j=L$ ({\sl i.e.,}~$\ell_{j}=1$) for $j=1,2$. 

We note that for ${j^\prime}\neq j$ the Eq.~(\ref{eq:generalCompet}) is linear in $\langle\nu\rangle_{j^\prime}$ if we assume steady state ($\dot{\nu}_{i,j}=0$). This allows us to express $\langle \nu \rangle_2$ as a function of $\langle \nu \rangle_1$:
\beq
\langle \nu \rangle_2=\frac{(\rho_1-\langle \nu \rangle_1)(1-\varphi\langle \nu \rangle_1)k_{+,1} - \langle \nu \rangle_1\left[k_{-,1}+p_1(1-\rho_1)\right]}{k_{+,1}\varphi(\rho_1 -\langle \nu \rangle_1)},
\eeq
assuming translation invariance. This equation can be inserted into Eq.~(\ref{eq:generalCompet}) for $j=2$, yielding an equation for $\langle \nu \rangle_1$ that can be solved numerically. With values for $\nu_1$ and $\nu_2$, we compute typical waiting time for TASEP $j=1$ as
\beq
\bar{\tau}_1 = \frac{1}{p_1} + \frac{\rho_1 - \langle\nu_1\rangle}{\rho_1k_{+,1}\left[1	- \varphi\left(	\langle\nu_1\rangle + \langle\nu_2\rangle	\right)\right]},
\eeq
which allows us to evaluate the current [Eq.~(\ref{eq:j1-periodic_general})].

The qualitative predictions of the model match the simulations well~(Fig.~\ref{fig:periodic_competition}b, right): we recover the independence of the two TASEPs and the high skew of current-density relations at $\varphi\ll1$, as well as the prominent shift of peak current to lower density at $\varphi\gg1$. While the model reproduces this feature it fails to faithfully reproduce absolute values for the current (see Fig.~\ref{fig:periodic_competition_supp}). The numerical exploration of this approximative framework for multiple TASEPs is possible, but beyond the scope of this article.

\subsection{Oscillations in kinetic rates and the propagation of fluctuations in coupled TASEPs}

In complex dynamical systems, coupling can do much more than just shift the steady-state operating point or induce a change in the average current-density relation. Pertinent examples include the emergent synchrony in phase-coupled oscillators or the propagation of temporal fluctuations through neural networks. Here we ask if the indirect coupling between two concurrently running TASEP processes through a shared token pool is strong enough to propagate (oscillatory) current fluctuations from one TASEP to the other~(Fig.~\ref{fig:isolated_oscillator}a).  To this end, we first explore the response of a single TASEP to an oscillation in one of its kinetic rates, the hopping rate $p$;~contrary to~\cite{zarai18} in which the number of lattices (mRNAs) oscillated with time and thus cause the effective hopping rate to oscillate.

\subsubsection{TASEP driven by an oscillating hopping rate}
\begin{figure}
\centering
   \includegraphics[width=8.6cm]{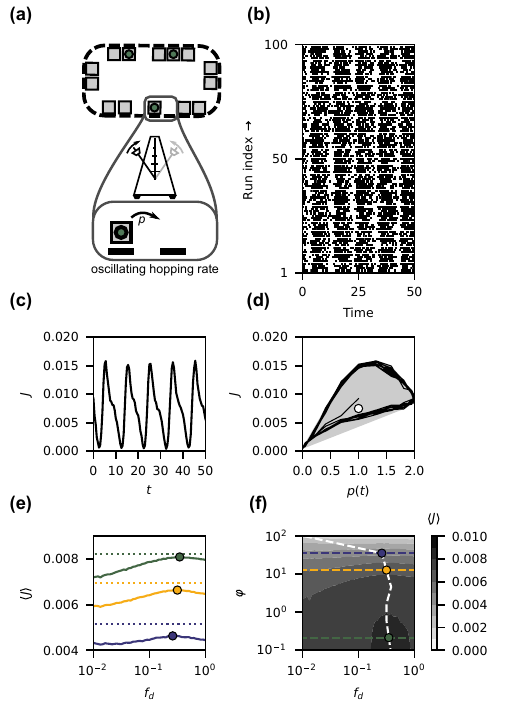}
   \caption{{\bf Response of a TASEP to an oscillatory drive.} (a)~Isolated TASEP with a harmonically oscillating hopping rate $p$. (b)~Typical ``hop'' profiles for an isolated TASEP. Each dot represents a hop. Note the existence of pulsing in hop occurrences. (c)~Current time trace averaged over $10^3$ runs. (d)~Phase portrait in the plane $p(t)$-$J$. Trajectory (black) demonstrates the existence of the hysteresis within a convex hull (gray). The trajectory starts in the middle of the convex hull due to initial equilibration. (e)~Response curves of average current as a function of driving frequency $f_d$ for different token scarcities $\varphi$ are non-monotonic, with maxima (filled circles) at $f_d\neq0$. Dotted lines denote the current in the absence of oscillations with matched average hopping rate. (f)~As in (e), but for multiple different values of $\varphi$. Colored dashed lines correspond to data shown in (e); white dashed line links the response maxima at different $\varphi$. Simulation data is for $\rho=1/10$, $k_+=0.1$, and $k_-=0.02$. For (b-d), we fixed $\varphi=4$ and $f_d=1/10$.}
   \label{fig:isolated_oscillator}
\end{figure}

Let the TASEP hopping rate be driven harmonically as $p=p_0\left(1-\sin2\pi f_d t\right)$, where $f_d$ is the driving frequency and $t$ is time~(Fig.~\ref{fig:isolated_oscillator}a). The time average $\langle p \rangle =p_0$, while the minimal and maximal value are $0$ and $2p_0$, respectively.
We distinguish two intuitive limits. In the adiabatic limit, the oscillations are slow ($f_d\ll1/\bar\tau$, where $\bar\tau$ is a typical particle waiting time) and the system will always remain close to the steady state supported by its parameters. In the noisy limit, the $p$ changes very rapidly and we hypothesize that the system settles to a steady state determined by the average hopping rate. 

In the adiabatic limit--where $J\left(p\left(t\right)\right)$ depends on the instantaneous hopping rate--the average current should be independent of the exact frequency of oscillations. To see this clearly, we first recall that the probability density function of hopping rates is $\eta(p)=1/\pi\left[{p_0^2-(p-p_0)^2}\right]^{1/2}$, symmetric around $p_0$ and independent of $f_d$, the same as for the classical oscillator.  To calculate the average current, we evaluate $\int_0^{2p_0}J_1(p)\eta(p) {\rm d }p$, avoiding any explicit time integration. Since the hopping rate enters into the current-density relations non-linearly~[Eqs.~(\ref{eq:nuFinite}) and~(\ref{eq:J1_finite})], the average current $\langle J\rangle\neq J(\langle p\rangle)$, as per Jensen inequality. Thus, in the adiabatic limit, the average current will be different from the one observed for $f_d=0$ even though the average hopping rate remains the same.

We recorded particle hop profiles time-locked to the oscillatory driving across replicate simulation runs~(Fig.~\ref{fig:isolated_oscillator}b). When averaged over multiple runs to suppress stochasticity (see Appendix~\ref{sec:simulations}), the oscillating current component emerges clearly~(Fig.~\ref{fig:isolated_oscillator}c). Interestingly, even though the hopping-rate oscillations are harmonic, the current response shows an increase that is steeper than the decrease. For slowly oscillating drive shown in Figs.~\ref{fig:isolated_oscillator}b and c, the current exhibits hysteresis in its phase portrait~(Fig.~\ref{fig:isolated_oscillator}d). This effect, likely a sign of deviation from true adiabacity, could be caused by token rearrangements during periods when hops are infrequent.

How does the time-averaged current depend on the driving frequency $f_d$ and on the token scarcity $\varphi$? First, irrespective of the driving frequency, the time-averaged current in a driven TASEP stays below the current of a non-driven TASEP with a matched average hopping rate (consistent with the adiabatic expectation; Fig.~\ref{fig:isolated_oscillator}e). Second, we observe an interesting behavior of the average current on the driving frequency: the current increases with frequency from the adiabatic limit towards the non-driven current value as the noisy limit is approached; but instead of increasing monotonically as expected, the current reaches a peak and then decreases for very high frequencies. The location and magnitude of this deviation from our intuitive expectation in the noisy limit depend on the degree of token scarcity $\varphi$~(Fig.~\ref{fig:isolated_oscillator}f), and are likely related to the increased importance of particle correlations on the lattice that our reasoning has neglected. Further, characteristic time $\tau$ [Eq.~(\ref{eq:tauFinite})] is not a good predictor for the of the peak frequency shown in Fig.~\ref{fig:isolated_oscillator}e,f. One could speculate that the $f_{\rm max}\approx 1/\tau$, but for $\varphi \approx 0.22, 12.92,$ and $35.94$ (colored lines in Fig.~\ref{fig:isolated_oscillator}), we obtain $1/\tau \approx 0.10, 0.09,$ and $0.07$, underestimating the true peak frequencies. We speculate that $f_{\rm max}$ is actually not directly related to the inverse of $\tau$, but is rather related to the (inverse) autocorrelation time of the system; this analysis, while interesting, is beyond the scope of the present paper.

\subsubsection{Propagation of oscillations in coupled TASEPs}
\begin{figure}
\centering
   \includegraphics[width=8.6cm]{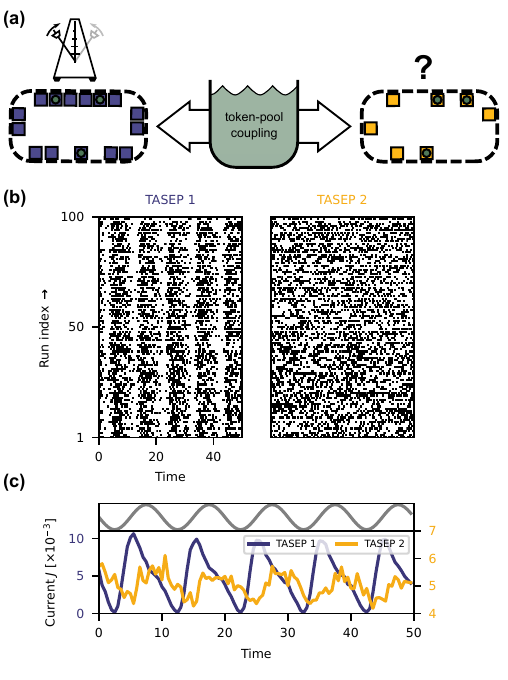}
   \caption{{\bf Current oscillations in the driven and the responding TASEP.} (a)~Two TASEPs share a common finite token pool. The hopping rate $p_1$ of the driven TASEP (left) oscillates in time. What is the current dynamics in the responding TASEP (right)? (b)~Typical hopping profiles, as in Fig.~\ref{fig:isolated_oscillator}b, here shown separately for the driven and the responding TASEP. Simulation data is for $\rho_1,\rho_2=0.5$, $\varphi=4$, and $f_d=1/10$. (c)~Top:~Oscillations in $p_1$ for guidance. Bottom:~Current time traces averaged over $10^3$ simulation runs, discretized at $\Delta t=1/2$. Colors match the TASEPs in (a); note different vertical scales for each trace. }
   \label{fig:periodic_oscillations1}
\end{figure}

We now consider a ``driven'' TASEP whose hopping rate $p_1$ is harmonically oscillating (as above), and which is sharing a finite token pool with another, ``responding'' TASEP~(Fig.~\ref{fig:periodic_oscillations1}a) with a constant hopping rate $p_2 = p_0$. We seek the frequency signature of harmonically oscillating $p_1$ in the current power spectrum of the responding TASEP.

Certain limits can be intuited. If tokens are abundant ($\varphi\sim0$), there is no coupling and oscillations cannot propagate from the driven to the responding TASEP. If tokens are very scarce ($\varphi\gg1$), the systems are totally dominated by stochastic token binding effects.  Between the two limits we expect a regime where oscillations can propagate efficiently. We also expect the individual densities of particles ($\rho_{1,2}$) to influence the degree of coupling. We explore these dependencies by simulations, where we let the two TASEPs reach steady state by simulating them initially at fixed hopping rates before turning on harmonic driving. 

Figure~\ref{fig:periodic_oscillations1}b shows typical hopping profiles of the driven and the responding TASEP. An average over simulation replicates clearly reveals that the current of the responding TASEP closely follows the oscillations in the hopping rate on the driven TASEP in anti-phase~(Fig.~\ref{fig:periodic_oscillations1}c), albeit with a smaller amplitude.

\begin{figure}
\centering
   \includegraphics[width=8.6cm]{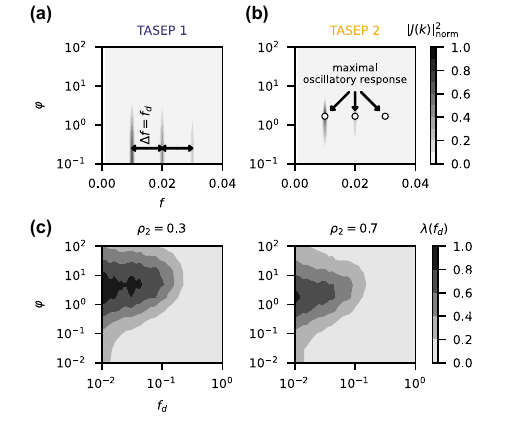}
   \caption{{\bf Spectral power of the driven and responding TASEPs, and fraction of explained variance.} (a) Driven TASEP exhibits discrete peaks corresponding to driving frequency $f_d$ and weaker overtones at multiples of $f_d$. This signature is strongest when tokens are abundant ($\varphi\rightarrow 0$) and decreases with token scarcity. (b) Responding TASEPs (right column) also exhibits strong spectrum components at these frequencies, but only within a window of token scarcities. In boths cases, $\rho_{1,2}=1/2$ and $f_d=1/10$.
   Spectral power is normalized by ${\rm max}_ {k\neq0} |J(k)|$.  Extremal responses are denoted if their value exceeds the average component for more than 2 standard deviations.
   (c)~The fraction of variance $\lambda(f_d)$ (grayscale) in the responding TASEP attributable to the influence by oscillatory driven TASEP, according to Eq.~(\ref{eq:partial_variance}). The plot is smoothed using a Gaussian filter (with smoothing parameter $\sigma_{\rm Gauss}=0.7$). For a given driving frequency $f_d$, we observe a non-monotonic dependence of $\lambda(f_d)$ on $\varphi$, with $\lambda$ peaking at a $\rho_2$-dependent value. Here, the density of particles on the driven TASEP has been fixed at $\rho_1=1/2$.
   }
   \label{fig:periodic_oscillations2}
\end{figure}

We next vary the driving frequency $f_d$ and the token scarcity $\varphi$.  Figure~\ref{fig:periodic_oscillations2} shows the power ($|J(k)|^2$) in different frequency components for the driven and the responding TASEP; the Fourier transform of the current is defined as $J(k)=\sum_{n=0}^{N-1}\exp\left(	-2\pi{\rm i} kn/N	\right)J(n)$ where $k$ denotes the discrete Fourier component, and $n$ is the sequential (time) index for the averaged current time trace. The driven TASEP contains a strong oscillatory component that monotonously decreases with increasing token scarcity~$\varphi$: when tokens are too scarce, stochastic token-particle binding events are rare and limiting enough so that a coherent current response is suppressed. The presence of higher harmonics in driven TASEP is expected since the current waveform deviates in shape from a perfect sinusoid~(Fig.~\ref{fig:isolated_oscillator}c).

The frequency response of the responding TASEP closely aligns with our  intuitive expectations:~for either $\varphi\to0$ or $\varphi\to \infty$, the oscillations are not transmitted from the driven TASEP, but the coupling is effective at intermediate values for $\varphi$, as revealed by excess spectral power at the driving frequency and its overtones.

To investigate the strength of the transmitted fluctuations, we compute the fraction of the current variance in the responding TASEP that is explained by Fourier components at the driving frequency and its multiples. Since biased variance \mbox{$\sigma^2\propto \sum_{k\neq0}|J(k)|^2$}, we define the degree of explained \emph{oscillation} variance as
\beq
\lambda(f_d)=\frac{\sum_{k_d}|J(k)|^2}{\sum_{k\neq0}|J(k)|^2},
\label{eq:partial_variance}
\eeq
where $k_d=f_d,2\times f_d,3\times f_d,\dots$ are multiples of the driving frequency. Because the driving frequency might not exactly correspond to a single discrete Fourier component due to sampling artifacts, the numerator of Eq.~(\ref{eq:partial_variance}) includes the closest discrete approximation to the driving frequency as well as the two nearest frequency components.

Figure~\ref{fig:periodic_oscillations2}c shows the efficiency of harmonic coupling between the two TASEPs as a function of token scarcity $\varphi$ and the driving frequency $f_d$, evaluated for two choices of particle density. The low driving frequencies transmit better (adiabatic limit), while the high frequency transmission through the shared token pool is damped.

The coupling also depends on the density of particles on TASEP lattices, because higher density of particles sequesters more tokens from the finite pool. We expect that if the densities of particles on the TASEPs are low and thus the fraction of free tokens is high, we should be closer to the infinite token pool scenario that does not couple the TASEPs effectively. In the other extreme, as the densities of particles on TASEPs tend to $1$, exclusion rather than token binding dominates the ability to hop, which should again limit the TASEP coupling. These heuristic considerations imply that the transmission of oscillations will be limited to the regime where (i)~the particle densities are sufficiently high to support significant removal of tokens from the pool yet not so high as to obstruct hopping; (ii)~token scarcity is sufficiently high such that the binding affects the free pool yet not so high that rare stochastic binding effects dominate behavior. These considerations are confirmed by simulations~(Fig.~\ref{fig:periodic_oscillations2}c), where we consequently  observe that higher particle densities move the maximal oscillatory response towards lower $\varphi$ and vice versa.

\begin{figure*}
\centering
   \includegraphics[width=17.2cm]{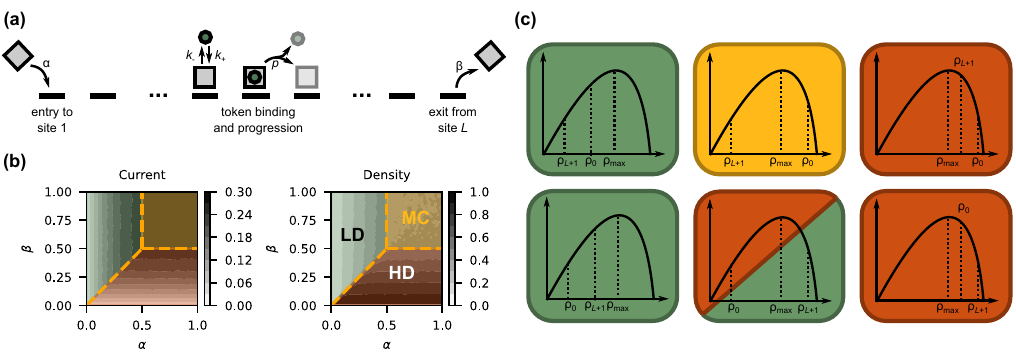}
   \caption{{\bf The basics of open-boundary TASEP.} (a)~Schematic. Particles enter with a rate $\alpha$, progress along the lattice [either in self-actuated or token-driven manner (shown here)], and exit with a rate $\beta$. (b)~Phase diagrams. Current and density as functions of rates $\alpha$ and $\beta$ for the standard TASEP. Colors denote different phases: red--high density, green--low density, and yellow--maximal current. (c)~Extremal-current principle for a current-density relation with a single maximum. The panel shows all different combinations of inequalities for boundary reservoirs densities ($\rho_{0},\rho_{L+1}$) and density supporting maximal current ($\rho_{\rm max}$). For most cases, the phase [denoted by color as in panel (b)] is independent of the exact current-density relation; for the case when $\rho_0<\rho_{\rm max}$ and $\rho_{\rm max}<\rho_{L+1}$, currents need to be calculated for boundary reservoir densities and the minimal value determines the phase.
   }
   \label{fig:open_basic}
\end{figure*}

Taken together, this and the preceding section demonstrate that even weak and rather indirect coupling through the pool of shared finite resources can propagate fluctuations (or signals) between two processes that we might otherwise be tempted to analyze independently.

\section{Open-boundary token-driven TASEP}
Previous section focused on systems with a fixed number of particles. By considering the system with open boundaries~(Fig.~\ref{fig:open_basic}a), the density ceases to be a control parameter, because it becomes dependent on the rates of entry ($\alpha$), hopping ($p$), and exit ($\beta$). 
In the standard formulation, the open-boundary TASEP exhibits three phases: low density, high density, and maximum current phase~(Fig.~\ref{fig:open_basic}b,c). In the low density (LD) phase, the entry rate determines the density, while in the high density (HD) phase, the current and the density are determined by the exit rate. In a maximum current (MC) phase the density becomes independent of entry and exit rates.

\subsection{Extremal current principle}
We construct the phase diagram of the open-boundary TASEP using an extremal principle~\cite{popkov99,krug91,shaw03}. This approach is useful as it connects the current-density relations of a closed system from previous sections to the open-boundary TASEP. We imagine that the open system connects two virtual reservoirs at the opposite ends of the lattice. These two reservoirs, one in front of the first site and one after the last site, have densities $\rho_0$ and $\rho_{L+1}$, respectively. While the hopping rate $p$ is kept the same for the particles exiting and entering the reservoirs, the densities $\rho_0$ and $\rho_{L+1}$ are tuned such that the particles enter and exit the system with rates $\alpha$ and $\beta$, respectively. 

In general, the current derived from the extremal principle is defined by extrema within the interval bounded by the densities of the boundary reservoirs~(Fig.~\ref{fig:open_basic}c):
\beq
J=
\begin{cases}
\max  J(\rho)&\text{for }  \rho_{L+1}<\rho<\rho_0\\
\min J(\rho) & \text{for } \rho_0<\rho<\rho_{L+1}
\label{eq:extremal}
\end{cases}.
\eeq
While the expression depends on the boundary reservoir densities $\rho_0$ and $\rho_{L+1}$, there is no direct way of fixing these densities in advance. A profound consequence of the extremal principle is that the phase diagram will retain the number of distinct phases if the number of extrema in the $J(\rho)$ stays the same. We can heuristically understand the extremal current principle as if the TASEP were trying to maximize the particle flow from the pool with the highest density towards the one with the lower. If the latter is on the side towards which the particles are moving, the current will be minimized within the interval defined by the densities of the boundary reservoirs. Below we use the extremal current principle to derive the phase diagram of the standard TASEP, as it will guide our subsequent derivation for the token-driven case. 

The continuity equation ($\nabla \cdot J=-\partial_t \rho$) in the steady state requires that $\nabla\cdot J=0$. Therefore, at the boundaries, this condition gives:
\begin{eqnarray}
0&=&\alpha\langle 1-\sigma_1\rangle-J_{1\to2}
\label{eq:boundary0}\\
0&=&J_{L-1\to L}-\beta\langle \sigma_L\rangle,
\label{eq:boundaryLp1}
\end{eqnarray}
for the entry and exit boundaries, respectively. Current-density relations established for TASEP with periodic boundary conditions imitate the current in open systems in the thermodynamic limit of large $L$. For the standard TASEP, we assume that $J_{i\to i+1}\approx p\rho_i(1-\rho_i)$ where we assumed $\rho_i\approx \rho_{i+1}$. Assuming this slow variation in density, we posit $\rho_0\approx\rho_1$ and $\rho_L\approx\rho_{L+1}$, which leads to $\rho_0=\alpha/p$ and $\rho_{L+1}=1-\beta/p$. These equalities allow us to use conditions from Eq.~(\ref{eq:extremal}) together with current-density relations to establish phase diagrams for $J$ and $\rho$ as a function of $\alpha$ and $\beta$. 

For standard TASEP, three regions (LD, HD, and MC; see above) constitute the phase diagram. When the entry rate $\alpha$ is above the exit rate $\beta$, and when $\beta<p/2$, high density regime occurs in which $\rho = 1 - \beta/p$ and $J=\beta(1-\beta)$. When the exit rate $\beta$ is higher than the entry rate $\alpha$, while $\alpha<p/2$, then the density increases linearly with $\alpha$, {\sl i.e.,}~$\rho =\alpha$, and current reads $J=\alpha(1-\alpha)$. The phase transition across the $\alpha=\beta$ line is discontinuous in density. However, when both entry and exit rates are above $p/2$, the current is maximal, {\sl i.e.,}~$J = p/4$, which is achieved at $\rho = 1/2$. Note, that in the thermodynamic limit of $L\to \infty$, such a phase diagram~(Fig.~\ref{fig:open_basic}b) is exact~\cite{derrida92,schuetz93,kolomeisky98}.

\subsection{Infinite pool of tokens}
\begin{figure}
\centering
   \includegraphics[width=8.6cm]{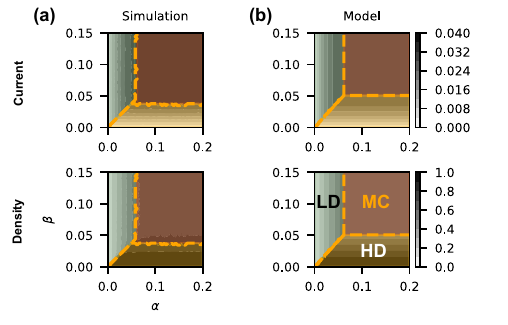}
   \caption{{\bf Phase diagrams for open-boundary TASEP with an infinite token pool.} (a)~Simulation results obtained for $k_+=0.1$ and $k_-=0.02$ over $50\times50$ grid. Phase color code is as in Fig.~\ref{fig:open_basic}. We estimated the phase boundaries between LD-HD and HD-MC phases by taking the first point at which density exceeds $\rho_{\rm max}$ at a given $\beta$ and first point at which current is equal to $J_{\rm max}=J(\rho_{\rm max})$ at fixed $\alpha$, respectively. (b) Same as for the (a), but for model results.}
   \label{fig:open_infinite_phase}
\end{figure}

We first analyze an open-boundary TASEP with an infinite pool of tokens. In simulations, we vary the entry and exit rates at fixed $k_+$ and $k_-$, to evaluate the average density as well as current~(Fig.~\ref{fig:open_infinite_phase}). We assume that the release of the particle at the end of the lattice does not require a token; we further assume that particles at the last site do not bind the token at all. Simulations show that, qualitatively, the token-driven TASEP preserves the number of phases as well as the general structure of the phase diagram when compared to the standard TASEP~(Fig.~\ref{fig:open_basic}). However, the maximal current and the corresponding density change, as we rationalize below.

We use the extremal current principle combined with the mean field approximation for the current-density relation~[Eqs.~(\ref{eq:averageNu}) and (\ref{eq:j0-periodic})]. Condition $\nabla\cdot J=0$ at both ends of the lattice leads to:
\begin{eqnarray}
\alpha( 1-\rho_0)&=&p\frac{k_+\rho_0}{k_- + p(1-\rho_0)+k_+} (1-\rho_0),\\
\beta\rho_{L+1}&=&p\frac{k_+\rho_{L+1}}{k_- + p(1-\rho_{L+1})+k_+} (1-\rho_{L+1}),
\end{eqnarray}
where we assumed again that $\rho_0\approx\rho_1$ and $\rho_L\approx\rho_{L+1}$ as well as that the steady-state result for $\langle \nu \rangle$ in~Eq.~(\ref{eq:averageNu}) is valid if we replace $\rho$ with $\rho_0$ or $\rho_{L+1}$, respectively. We further mathematized the fact that the tokens do not bind to the particle at the site $L$ and thus release is proportional to the $\rho_L$.

Boundary reservoirs consequently have the following densities:
\begin{eqnarray}
\rho_0&=&\frac{\alpha\left(k_++p+k_-\right)}{p\left(k_++\alpha\right)},\\
\rho_{L+1}&=&\frac{pk_+-\beta\left(	k_++p+k_-	\right)}{p\left(k_+-\beta	\right)}\label{eq:rhoLp1}.
\end{eqnarray}
As a sanity check, we note that as $k_+\to\infty$, we obtain $\rho_0=\alpha/p$ and $\rho_{L+1}=1-\beta/p$, the reservoir densities of the standard TASEP. Together with the expression for $\rho_{\rm max}$~[Eq.~(\ref{eq:maxCurrent_infinite})], we can now estimate different regions of the phase diagram, at least for $k_+\gg\beta$. Unfortunately, Eq.~(\ref{eq:rhoLp1}) has a first-order pole at $\beta=k_+$, suggesting that the phase behavior could be unstable, since the boundary reservoir density $\rho_{L+1}$ changes sign when $\beta$ passes through the pole. 

To deal with this inconvenience, we turn our attention to the refined model for the current [Eq.~(\ref{eq:j1-periodic})], which we combine with  Eqs.~(\ref{eq:boundary0}) and~(\ref{eq:boundaryLp1}) to get the boundary densities $\rho_0$ and $\rho_{L+1}$:
\begin{eqnarray}
\rho_0&=&\frac{k_+ p (\alpha+\hat{k})+p^2 (\alpha+k_+)}{{2 k_+ p^2}}-\\
&-&\frac{\sqrt{p^2 \left[\delta^2-4 \alpha k_+ \left(p \hat{k}+k_+ \hat{k}+p^2\right)\right]}}{2 k_+ p^2},\nonumber\\
\rho_{L+1}&=&-\frac{\beta (k_+ + p) - k_+ (\hat{k} + 2 p)}{2 k_+ p}+\label{eq:rhoLp1Ref}\\
&+&\frac{\sqrt{
 k_+^2 (\beta + \hat{k})^2 + 2 \beta (\beta + \hat{k}) k_+ p + 
   \beta^2 p^2}}{2 k_+ p},\nonumber
\end{eqnarray}
where $\hat{k}=k_++k_-$ and $\delta=\alpha (k_++p)+k_+ (\hat{k}+p)$.  Equation~(\ref{eq:rhoLp1Ref}), in contrast to Eq.~(\ref{eq:rhoLp1}), does not have any poles when $\beta$ increases. Together with the extremal current principle, it provides us with a broadly applicable approximation for the open-boundary TASEP which we explore in the rest of the section.

We first check that the reservoir density expressions recover $\rho_0=\alpha/p$ and $\rho_{L+1}=1-\beta/p$ in the limit $k_+\to\infty$, as expected for the standard TASEP. It is also possible to compute a closed-form expression for the density resulting in maximal current, providing all the necessary ingredients to evaluate the phase diagram. Comparison with simulations reveals a very good match~(Fig.~\ref{fig:open_infinite_phase}).

\begin{figure}
\centering
   \includegraphics[width=8.6cm]{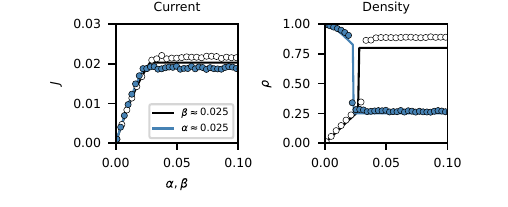}
   \caption{{\bf Current and density as a function of entry and exit rates in an open-boundary TASEP with an infinite token pool.} (Left) Current at fixed $\alpha=0.025$ (blue disks) as a function of $\beta$, or at fixed $\beta=0.025$ as a function of $\alpha$ (white disks) for $k_+=0.1$ and $k_-=0.02$. Current increases continuously with both parameters. (Right) As in the left panel, but for the density. While the current changes continuously, the density undergoes an abrupt change. These dependencies are correctly captured by the model (solid lines). Simulation data is from Fig.~\ref{fig:open_infinite_phase}.
   }
   \label{fig:open_infinite_sections}
\end{figure}

Our analytical approximations correctly predict  the continuous change in the current as well as abrupt jumps in density as $\alpha$ or $\beta$ increase~(Fig.~\ref{fig:open_infinite_sections}). 
Specifically, the approximations correctly predict transition lines at which the phase changes as a function of entry or exit rates, confirming that the refined mean-field theory can explain parameter regimes supporting each of the three accessible phases.
\subsection{Finite pool of tokens}
A finite token pool had a far-reaching impact on the current-density relation for the system with periodic boundary conditions: maximal current was reached at lower densities compared to the standard TASEP as token scarcity increased, $\varphi\gg1$. With the extremal current principle and the current-density relations [Eq.~(\ref{eq:J1_finite})], we can predict the structure of the phase diagram for the token-driven TASEP with a finite token pool.

\begin{figure}
\centering
   \includegraphics[width=8.6cm]{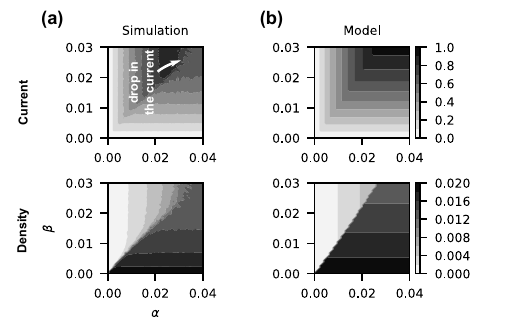}
   \caption{{\bf Current and density for open-boundary TASEP with a finite token pool.} Current (top row) and density (bottom row) are shown in gray scale as a function of entry and exist rates, $\alpha$ and $\beta$. (a)~Simulation and (b)~model results, respectively, for $\varphi\approx2.51$. The model fails to predict the abrupt drop in the current (arrow) when $\alpha$ passes a critical value for a given $\beta$. Results for $k_+=0.1$ and $k_-=0.02$, evaluated over $50\times50$ grid. 
   }
   \label{fig:open_finite_Phi}
\end{figure}

We simulate TASEP for different values of token scarcity, $\varphi$. For $\varphi<1$, the model correctly describes both current and density as functions of $\alpha$ and $\beta$ and retains the structure largely similar to the one for the infinite tokens case. Yet, when $\varphi>1$, the model fails to correctly describe the current~(Fig.~\ref{fig:open_finite_Phi}). The current does not increase monotonically with $\alpha$ at a given $\beta$ (as it does for an open TASEP for $\varphi\to0$), but rather experiences a sharp decrease from a maximum towards a plateau at a lower value.

The abrupt change in current becomes even more apparent when we study cross-sections at fixed $\beta$ for different $\varphi$~(Appendix~\ref{sec:detailed_open}, Fig.~\ref{fig:open_finite_sections}). While the model correctly predicts $\alpha$ at which the phase transition occurs, if fails to predict the lower current plateau whose height depends on $\beta$. Additionally, the model predicts monotonic changes in current, which only seems to hold for $\varphi<1$, but not for $\varphi>1$--the model does not predict the drop in the current around phase transition line~(Fig.~\ref{fig:open_finite_Phi}a). The deviations likely originate due to the long-reaching correlations within the dense traffic in the $\alpha$-independent phase, which the mean-field does not capture properly.

\subsection{Spontaneous particle drop-off}
\begin{figure}
\centering
   \includegraphics[width=8.6cm]{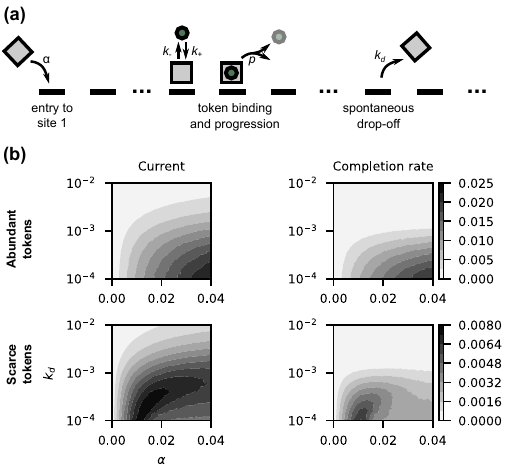}
   \caption{{\bf Spontaneous particle drop-off in token-driven open TASEPs.} (a)~Schematic of a token-driven open-boundary TASEP with spontaneous particle drop-off. Particles without a token drop from the lattice with characteristic rate $k_d$. (b)~Current and completion rate as a function of drop-off rate $k_d$ and entry rate $\alpha$ for abundant ($\varphi=0.01$, top row) or scarce ($\varphi\approx15.8$, bottom row) tokens. Simulation is for $k_+=0.1$ and $k_-=0.02$ over $50\times50$ grid. To avoid the congestion at the end of the lattice, we set $\beta=100$.
    }
   \label{fig:open_dropoff}
\end{figure}

To expand on the token-driven open-boundary TASEP framework, we  consider the case where the unbound particles can spontaneously drop off the lattice with rate $k_d$~(Fig.~\ref{fig:open_dropoff}a). This setting is a limiting case of the model where particles enter and exit the lattice at its ends, as well as attach and detach in its bulk~\cite{parmeggiani03}. Reference~\cite{parmeggiani03} demonstrates that such intra-lattice exchange of particles supports non-trivial density profiles, where high- and low-density regions coexist and are separated by transition region, similar to  density phases coexistence in systems with local defects~(Ref.~\cite{janowsky92} and Sec.~\ref{sec:slow}). Bulk particle exchange rates differ from other rates discussed previously, because the impact of bulk particle exchange processes will depend on the system size $L$.

Here, we focus on the case where particles enter the lattice only at the beginning but can detach either in the  bulk or at the end. A moderate amount of ``drop-off'' ({\sl i.e.,}~detachment in the bulk of the lattice) could reduce the density on the lattice, thus improving the utilization of the finite token pool and counteracting traffic jams due to stuck particles waiting for their tokens.  Clearly, excessively high drop-off rates (or a very long lattice, {\sl i.e.,} $L\to \infty$ at constant $k_d$) would completely decimate the likelihood of particles ever reaching the end of the lattice, so we envision the existence of an optimal drop-off rate that maximizes the current. This is what we explore next.

We set $\beta\gg p,\alpha,k_d$ to reduce the likelihood of congestion at the end of the lattice. We use simulations to compute the current and the completion rate~(Fig.~\ref{fig:open_dropoff}b). Since the density varies across the lattice, the current also changes from site to site~(see Appendix~\ref{sec:continuous}). Therefore, we define current here as an average across the whole lattice, whereas the completion rate is the particle current from the last lattice site (typically lower than the average current). For abundant tokens, we observe that spontaneous drop-off has mild effects unless $k_d$ is increased sufficiently, after which the current and completion rate gradually decrease. For scarce tokens, in contrast, the current and completion rate exhibit a non-monotonic behavior as $k_d$ increases at a fixed $\alpha$: there exists an optimal $k_d(\alpha)$ that maximizes the current and the completion rate, as we expected.  We further show that if the drop-off parameter $k_d$ is scaled inversely with the lattice length $L$, the results shown in Fig.~\ref{fig:open_dropoff} remain robust and are representative of the thermodynamic limit $L\to \infty$ (see Appendix~\ref{sec:length}).

\section{Discussion}
We performed an extensive study of token-driven TASEP and its many variants. We used a combination of mean-field models to establish intuitions and limits, and simulations to compute exact results. Here we highlight several interesting links between the studied TASEP variants and the experimentally observed systems.

\emph{Real and optimal regimes of TASEP function.} Stochasticity intrinsic to TASEP means that TASEPs generally do not reach the maximal current achievable by particles hopping in perfect synchrony. This holds true for the standard as well as the token-driven TASEP. It is interesting to speculate whether there exist passive or active driving and control mechanisms that would approach such maximal possible current.  These mechanisms could either be agnostic about the global state of the TASEP, as with periodic driving, or could include feedback that would couple any controlled TASEP rate ({\sl e.g.}, of hopping) to the instantaneous current. While periodic driving of the hopping rate that we explored in this paper yielded currents strictly lower than the equivalent constant hopping rate, this need not be the case for other driving protocols. A broad open  question of biological relevance would thus concern plausible schemes to maximize the current at finite TASEP resources: by means of time-dependent driving to synchronize particle hops, by means of dynamic entry rate control to prevent traffic jams, or by means of spontaneous particle drop-off to dissolve token-sequestering accumulation of particles on the lattice. Alternatively, instead of asking ``What is optimal?'' we could ask ``Where in the TASEP phase diagram do we find natural systems?'' Cells, for instance, react to environmental changes by altering the abundance of components involved in protein synthesis~\cite{scott10,dai16,bremer96}; concretely, bacteria regulate the quantities of translation factors and ribosomes in synchrony~\cite{dai16,maaloe79,gordon70,blumenthal76,furano76}.  What are the consequences of such regulatory and feedback schemes? Are they an indication that evolution optimized a particular aspect of TASEP performance, and if so, which one? Much remains to be done at the interface between optimization approaches and inference from data~\cite{mlynarski21}, especially in biological TASEP contexts.

\emph{Details of token-particle interactions} matter when tokens are scarce. Current can be substantially different if we consider the baseline scenario, or its refractory or token-retention modifications. For instance, during the elongation cycle of translation, ribosomes pass through several sub-steps, only one of which (translocation) involves the movement of the ribosome by a single site~\cite{rodnina18}. This is because, prior to translocation, decoding and peptide bond formation must occur, which together constitute a ``refractory period.'' Molecular motors, on the other hand, require ATP hydrolysis (into ADP) to occur, before ADP can detach and be recycled. Here, we can consider ATP/ADP to be the required token which exists in either the ``loded'' (ATP) or the ``unloaded'' (ADP) states. The molecular motor is therefore an example of the token-retention scenario~\cite{alberts02}. Further everyday life examples abound. During software development, a developer ({\sl i.e.,}~a token) undertakes a task ({\sl i.e.,}~a particle waiting to hop forward) to submit a solution, which requires approval by the supervisor before going into production~({\sl i.e.,}~a particle waiting in a refractory period). Meanwhile, the developer can already work on another task. Alternatively, a bureaucrat (again, a token) awards themselves a coffee after each processed application (a particle). The bureaucrat thus remains unavailable ({\sl i.e.,} token-retention state) for starting a new application, thereby increasing the overall completion times. Our analysis suggests that in all these micro- or macroscopic cases, it is beneficial to release the tokens from the particle as soon as possible when tokens are limiting, even if that means that the task must temporarily remain in the refractory state.

\emph{Further examples of TASEP extensions.} First, we considered the impact of a  single slow site on the token-driven TASEP density profile and its current. In cells, ribosomes may encounter transcripts that contain rare codons--{\sl i.e.}, codons whose complementary tRNA is present at low concentrations--making the passage of the ribosome across this ``barier'' slow~\cite{hanson17}. In vehicular or pedestrian traffic, a crossroad, a traffic light, or a tollbooth can cause a similar bottleneck that leads to the build-up of traffic congestion in front of the barrier. Second, bacterial cells also provide an example of how $10^2-10^4$ simultaneously-translating ribosomes can become coupled as they compete for the limited pool of shared translation factors~\cite{bartholomaeus16}. Third, the messenger RNAs (mRNAs) in the bacterium {\sl E.~coli} can also be interpreted as competing for a pool of $5-75\times10^4$ particles (ribosomes)~\cite{bremer96}, only a fraction of which are available for binding while the rest are busy with translation. This consideration has been explored by~Greulich {\sl et al.}~\cite{greulich12}, who showed that the common pool of particles couples TASEPs through initiation rates that depend on the availability of free particles. A unified, fully-stochastic framework (going beyond the mean-field~\cite{kavcic20}) which would combine token-driven open TASEP with finite pools of tokens as well as particles would elucidate how all these constraints jointly shape the TASEP dynamics.

\emph{Consequences of abrupt current changes in open TASEP.} In case of finite token pool for an open-boundary TASEP, the model predictions deviate from the observed abrupt drop in current upon crossing the critical value for the entry rate~(Figs.~\ref{fig:open_finite_Phi},\ref{fig:open_finite_sections}). It is interesting to speculate about the impact of such abrupt current changes for naturally occurring token-driven TASEPs. For ribosomes engaged in protein synthesis, where $\alpha$ corresponds to the regulated rate of initiation, such an abrupt drop could pose a challenge. Our analysis suggests that a regulatory scheme based on controlling $\alpha$ would either lose regulatory power as $\alpha$ approached the transition point, or could lead to the paradoxical situation where the current would decrease with increasing $\alpha$. A more involved analysis (Appendix~\ref{sec:TwoStageOpen}) demonstrates that it is possible to suppress the development of abrupt current drops, but can only at the expense of lower maximal current.

\emph{Extended particles.} Several extensions of the framework would further improve its relevance for the understanding of experimentally accessible systems. The token-driven movement of ribosomes during protein synthesis proceeds in discrete steps while the ribosome footprint extends over $\approx8.33$ sites on the mRNA ``lattice''~\cite{mohammad19,woolstenhulme15}. Mean-field descriptions of this process~\cite{shaw03,lakatos03,kavcic20} should be complemented by a detailed exact or simulation-based treatment.

\emph{Consequences of observed TASEP heterogeneity.} Quantitative contact with data will require us to depart from the stylized assumptions of toy models and work out the regime of highly heterogenous TASEPs. This naturally leads towards a statistical mechanics of TASEP ensembles that is very relevant for biological systems. For example,  different mRNAs in living cells have different initiation rates determined by the biophysical details of ribosome recruitment~\cite{salis09}. With the availability of detailed data describing the whole distribution of ribosomes across different mRNAs, these rates of initiation and elongation can be inferred~\cite{szavitsnossan20b}. How such inferred heterogeneity interacts with the dynamics of TASEPs coupled through a shared token pool remains to be understood.

\emph{Experimental tests.} Applications of TASEP  to biological systems are well suited for experimental verification. The reconstituted {\sl in vitro} systems of biopolymer synthesis ({\sl e.g.,} transcription-translation systems) are increasingly advanced and well-controlled in terms of their constituents. These advances enable targeted experiments to assess the effects of tokens on TASEP dynamics, specifically, to relate our theoretical framework to defined biochemical parameters. For example, by titrating the abundance of translation factors and measuring the resulting rate of protein synthesis and the density of ribosome traffic, we could directly relate our theoretical results to  experimental data. Such {\sl in vitro} experiments would directly connect to studies in which translation factors are titrated {\sl in vivo}, either individually~\cite{kavcic20,cole87,olsson96} or in combination~\cite{kavcic20}, to isolate the role of natural physiological and cellular context.

\acknowledgements
B.K.~thanks Stefano Elefante, Simon Rella, and Michal Hled\' ik for their help with the usage of the cluster. B.K.~additionally thanks C\u alin Guet and his group for help and advice. We thank M.~Hennessey-Wesen and Luca Ciandrini for constructive comments on the manuscript. We thank Ankita Gupta (Indian Institute of Technology) for spotting a typographical error in Eq.~(\ref{eq:rhoLp1Ref}) in the preprint version of this paper.

\appendix
\setcounter{table}{0}
\renewcommand{\thetable}{A\arabic{table}}%
\setcounter{figure}{0}
\renewcommand{\thefigure}{\thesection\arabic{figure}}%

\section{Details of stochastic simulations\label{sec:simulations}}
We implemented a Gillespie algorithm~\cite{gillespie77} to simulate TASEP systems. For a given state of the system, the algorithm computes all kinetic rates, $r_i$. Summing-up these rates yields a total rate $R=\sum_i r_i$, from which we calculate the time after which an event will happen. This time is given as $\tau=-\log(\xi)/R$, where $\xi$ is a random number drawn from a uniform distribution between $0$ and $1$. The probability for $i$-th event to transpire is weighted as $w_i=r_i/R$. This scheme allows for an efficient and exact simulation of stochastic processes.

Baseline results reported in this paper are computed for the lattice length $L=200$. For specific cases, we examined the effects of the lattice size (see Appendix~\ref{sec:length}). We deployed simulations to run in parallel on a computer cluster, and allow simulations to run until $10^6$ particle hops or $10^8$ iterations (occurrence of any event) transpired, whichever happened first. For periodic boundary conditions, we distributed particles randomly (without bound tokens) onto the lattice at the initiation. In an open boundary case, we initiated simulations with empty lattices, and let the simulation run until first particle reached the end of the lattice. We let the simulations run until $2.5\times 10^7$ events transpired before we started recording the density and counting forward hops.

Due to non-uniform discretization of time arising from Gillespie algorithm, we calculated--when applicable--averages weighted by the length of time intervals between iterations. To evaluate the current, we counted all forward hops and divided the total by the elapsed time. We normalized the count by the length of the lattice. 

For studies of oscillatory behavior, we averaged the current time trace over $10^3$ runs. We initially let the TASEPs equilibrate while keeping $p_1=p_2$. After $10^3$ steps, hopping rate $p_1$ began changing with time. We simplified this part of simulation by assuming that $\tau<1/f_d$, {\sl i.e.,} oscillations in $p_1$ are much slower than a typical time between events. This way $p_1\approx p_2\left(1+\sin f_d T_k\right)$, where $T_k=\sum_{i=0}^{k-1}\tau_i$ for $k$-th step and $\tau_i$ is a Gillespie-calculated event time. To fulfill Nyquist criterion, we calculated running average of the current over intervals $\Delta T=1/2f_{\max}$, where $f_{\rm max}$ is the highest oscillation frequency. We concluded each simulation run when time exceeded $10^3$. This time suffices that even the slowest oscillations undergoes $100$ periods. We normalized all currents per lattice site. 

We programmed simulations and analyzed the simulation results in Python (3.8.5), using libraries numpy (1.19.2),  scipy (1.5), and matplotlib (3.4.2), while we used Mathematica (11.3) for analytical calculations. We will deposit simulation results data to IST Austria repository for free access after acceptance of the article.

\section{Impact of lattice size on the current-density relations\label{sec:length}}
We further verified the impact of the lattice size on the current-density relations. As examples, we chose the cases with periodic boundary conditions with infinite and finite tokens. We further chose the case of open TASEP with spontaneous drop-off as an example in which we expect the lattice length to profoundly affect the dynamics. In the former cases we left the rates the same for all $L$, while in the latter we rescaled the drop-off rate by length. Specifically, this required $\tilde{k}_d= k_d \times (200/L)$, where 200 is the baseline lattice length used to report the main results in this paper.  

For all cases, we scaled the sampling with the length of the lattice ({\sl e.g.,} for twice-longer lattice, twice as many equilibration iterations, particle hops/iterations are required). We checked the lengths $L=100, 200, 400, 800$, and $1600$; the latter was omitted in the open-boundary case, since the computational time was too excessive for  parameter exploration. 

\begin{figure}
\centering
   \includegraphics[width=8.6cm]{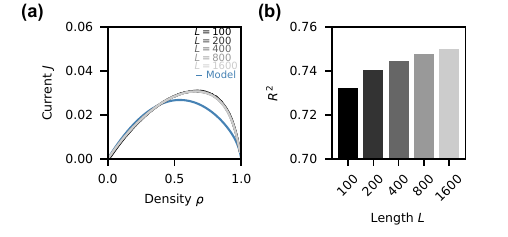}
   \caption{{\bf Current-density relations for token-driven TASEP on a ring in an infinite token pool for different lattices sizes.} (a)~Plot as in Fig.~\ref{fig:rho-J-infinite}, but for different sizes of the lattice; markers were omitted for clarity. Blue line denotes a prediction from a refined model~[Eq.~(\ref{eq:j1-periodic})] Lattice size $L$ has no significant effect on the current-density relations. (b)~Coefficient of determination $R^2$ calculated for simulation results compared to the model predictions. Data in the plots is for $k_+ = 0.1$ and $k_-=0.02$. 
    }
   \label{fig:rho-J-infinite_varlen}
\end{figure}

\begin{figure}
\centering
   \includegraphics[width=8.6cm]{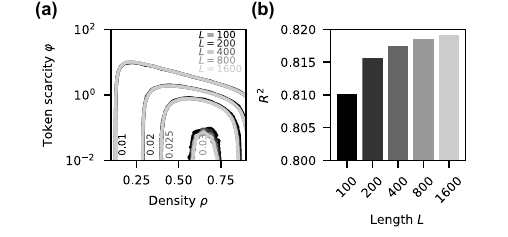}
   \caption{{\bf Effects of different lattice lengths for the case of periodic TASEP driven by finite tokens.} We varied the lattice length to check for the effects of $L$ on current-density relations for finite tokens. (a)~We show four current contours with indicated values; their color indicates the chosen length. Used $L$ for each case is indicated and color-coded. The resultant contours are largely independent of the length. (b)~Coefficient of determination $R^2$ calculated for simulation results compared to the model predictions. Both model and simulation results were evaluated on a $20 \times 100$ grid. Data in the plots is for $k_+ = 0.1$ and $k_-=0.02$.
   }
   \label{fig:periodic_finite_varlen}
\end{figure}

Figure~\ref{fig:rho-J-infinite_varlen} shows that current-density relations for periodic TASEP with infinite tokens. To compare the deviation of the simulation results from the model predictions, we evaluated the coefficient of determination $R^2 = 1 - \left[\sum_i (y_i - f_i)^2\right]/\left[\sum_i (y_i - \bar{y})^2\right]$, where $f_i$, $y_i$, and $\bar{y}$ are the current values obtained in model predictions, simulations, and a simulation average $\bar{y} = (1/N)\sum_i y_i$, respectively. The results are largely independent of the lattice length, with deviations from the model that show no strong lattice length trend. We further evaluated $R^2$ for the simulation results for the shortest ($L=100$) and  the longest ($L=1600$) lattice for the case of the periodic TASEP with infinite tokes, which yields $R^2 \approx 0.998$. As such, we can consider results obtained for $L=200$ throughout the article to be an adequate representation of the thermodynamic limit $L\to \infty$.
Likewise, as shown on Fig.~\ref{fig:periodic_finite_varlen}, current dependence on density and token scarcity remains largely independent of $L$ for the case of finite tokens.

\begin{figure}
\centering
   \includegraphics[width=8.6cm]{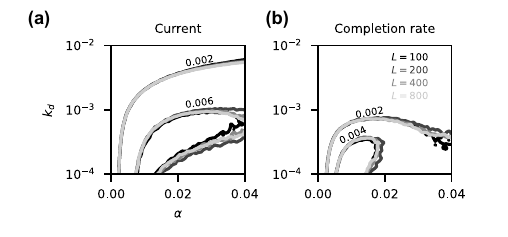}
   \caption{{\bf Spontaneous particle drop-off in token-driven open TASEPs of different lengths.} As Fig.~\ref{fig:open_dropoff}, but for different lengths and only two contours are shown. For the shortest lattice ($L=100$), there are differences in the shape of the current and completion-rate surfaces when compared to longer lattices. As described in the main text, $k_d$ was rescaled such that the drop-off rate $\tilde{k}_d= k_d \times (200/L)$ remained constant for a given lattice length. Simulations are for $k_+=0.1$ and $k_-=0.02$, $\varphi\approx15.8$, and evaluated over $50\times50$ grid. As is Fig.~\ref{fig:open_dropoff}, we set $\beta=100$ to avoid the congestion at the end of the lattice.
    }
   \label{fig:open_dropoff_varlen}
\end{figure}

The length has more impact on the case in which the lattice can lose unbound particles with characteristic rate $k_d$. As discussed above, we needed to rescale the drop-off parameter with length, similarly as discussed in Ref.~\cite{parmeggiani03}. Using this rescaling, we obtain comparable results for all chosen lengths, with minimal deviation at $L=100$ (Fig.~\ref{fig:open_dropoff_varlen}).

\section{Density-current cross-sections for periodic TASEP\label{sec:density-current_relations}}
In this Section we show complementary density-current relations for data shown in colormap plots.
\begin{figure}
\centering
   \includegraphics[width=8.6cm]{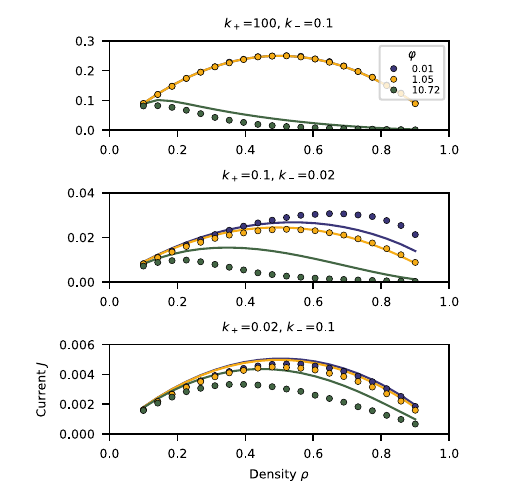}
   \caption{{\bf Current as a function of density for the case of periodic lattice and finite tokens.}  Figure shows current-density relations for a few representative values of token scarcity ($\varphi$, see legend) for data show in Fig.~\ref{fig:periodic_finite_surfaces}. Solid lines represent model predictions and markers denote simulation results.}
   \label{fig:periodic_finite_density-rho_examples_plot}
\end{figure}

\begin{figure}
\centering
   \includegraphics[width=8.6cm]{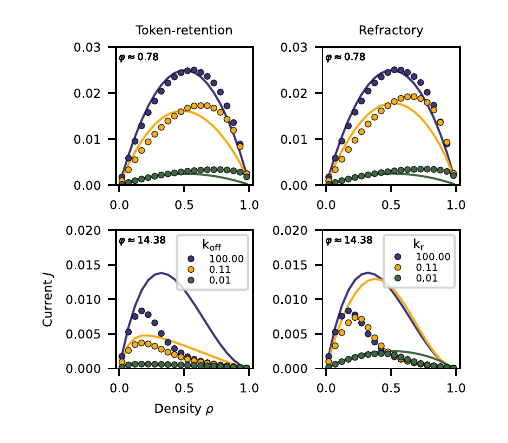}
   \caption{{\bf Current as a function of density for the case of periodic lattice, finite tokens and two-stage progression of hopping particles.}  Figure shows current-density relations for a few representative values of token scarcity ($\varphi$, see legend) for data shown in Fig.~\ref{fig:twostage}. Solid lines represent model predictions and markers denote simulation results.}
   \label{fig:twostage_supp}
\end{figure}

\begin{figure}
\centering
   \includegraphics[width=8.6cm]{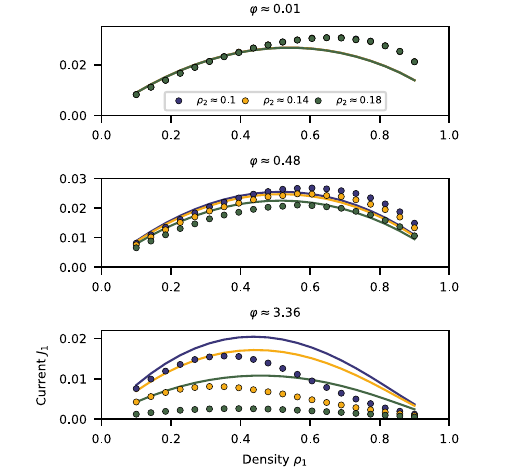}
   \caption{{\bf Current as a function of density for the case of two competing periodic TASEPs with finite tokens.}  Figure shows current-density relations for a few representative values of token scarcity ($\varphi$, see titles) for data shown in Fig.~\ref{fig:periodic_competition} and for different particle densities of the competing TASEP. Solid lines represent model predictions and markers denote simulation results.}
   \label{fig:periodic_competition_supp}
\end{figure}

\section{Current and density for open TASEP as a function of $\alpha,\beta$ obtained in simulations and in modeling\label{sec:detailed_open}}
\begin{figure}
\centering
   \includegraphics[width=8.6cm]{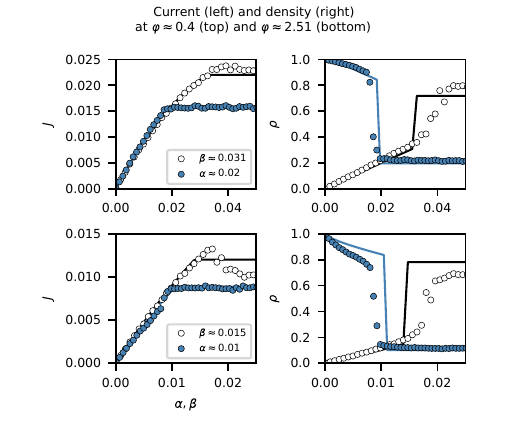}
   \caption{{\bf Current and density as a function of $\alpha,\beta$ as obtained in simulations and in modeling.} Top:~Simulation and model results for $\varphi\approx0.4$. Left: Current at fixed $\alpha$ (blue disks) or $\beta$ (white disks); see legend for values of fixed quantities. Note a continuous increase in current with increasing $\beta$ and $\alpha$, respectively. Solid lines correspond to model predictions. Right: As in the left panel, but for the density. Bottom: as for Top panels, but for $\varphi=2.51$. Simulation data is from Fig.~\ref{fig:open_finite_Phi}
   }
   \label{fig:open_finite_sections}
\end{figure}

Figure~\ref{fig:open_finite_sections} shows the details of the abrupt transition in current when critical $\alpha$ is exceeded at a fixed $\beta$.
For $\varphi<1$, current changes continuously while the density undergoes an abrupt change that the model correctly predicts (as in Fig.~\ref{fig:open_infinite_sections}). For $\varphi>1$, the model is unable to predict the abrupt drop in current upon transiting into the $\alpha$-independent regime.

\section{Token-retaining and refractory states in open-boundary TASEP\label{sec:TwoStageOpen}}

\begin{figure}[t!]
\centering
   \includegraphics[width=8.6cm]{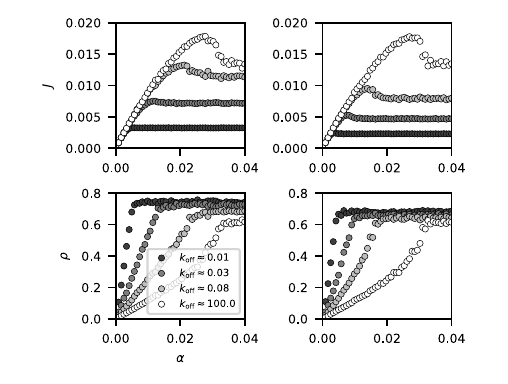}
   \caption{{\bf Current and density as a function of $\alpha$ obtained in simulations at $\beta=0.03$ and $\varphi=2.51$.} Top:~Current for particles with refractory (left) and token-retaining (right) state. Note that for $k_r,k_{\rm off} \gg k_+,k_-,p$ (white circles)  we obtain the same bump in current relation. When $k_r,k_{\rm off}$ are reduced (see legend), the bump becomes slightly reduced at the expense of lower maximal current. Bottom: as on the top, but for density.}
   \label{fig:open_finite_twoStages}
\end{figure}

The abrupt drop in current upon crossing the critical value of $\alpha$ at fixed $\beta$ was a surprising observation not captured by the mean-field model. Would this drop persist if particles entered either refractory or token-retaining state upon hopping forward? In refractory state, particles do not take up another token and thus do not reduce the number of particles taking up tokens. On the other hand, retaining particles for prolonged time upon forward hop could drain the token pool. Here, we computationally study how varying $k_r$ and $k_{\rm off}$ impacts the current transition between low-density and $\alpha$-independent regime~(Fig.~\ref{fig:open_finite_twoStages}).

We observe in simulations that while for particles with either refractory or token-retaining state bump disappears, former requires less reduction in $k_r$ compared to the latter's $k_{\rm off}$. However, lowering $k_{\rm off}$ also leads to slightly elevated particle density. In summary, either of these effects recovers monotonically increasing current dependency on $\alpha$.

\section{Site-dependent density and current--from discrete lattice to continuous variation\label{sec:continuous}}

Including spontaneous drop-off alters the density profile; while the density profile remains time-independent ($\partial_t \rho=0$), it changes along the lattice. Below we suggest a framework for the analysis of such site-dependent density and current profiles. For the case with a spontaneous drop-off, from the continuity equation with sink term ($\nabla\cdot J=-\partial_t \rho-S$, where $S$ is the sink rate) we arrive to the difference equation
\beq
J_{i-1}-J_i=k_d\sigma_i\label{eq:sink}.
\eeq
This equation accounts for the rate of particle drop-off ($k_d\sigma_i$). Taking the sink term into account, we rewrite the differential equations for token-free particles:
\beq
\dot{\sigma_i}=k_-\nu_i +\nu_{i-1}(1-\rho_i)p -\sigma_i\left[k_+\left(1-\varphi \langle\nu\rangle\right) +k_d\right]\label{eq:sigmaWsink}.
\eeq
Equation~(\ref{eq:tokens_finite}) for $\nu_i$ remains unchanged. In this system of equations [Eqs.~(\ref{eq:tokens_finite}), (\ref{eq:sink}), and (\ref{eq:sigmaWsink})], we cannot assume the translation invariance of density profile ($\rho_i\neq\rho_j$ for $j\neq i$). Thus, solving this system requires finding explicit expressions for site-dependency of $\rho_i$, $\nu_i$, and $\sigma_i$, which is likely to depend on length of the lattice. 

One potential simplification of the system is through replacing the discretized lattice with its continuous analogue. This way, we assume a continuous variation in $\rho$, $\nu$, and $\sigma$. Concretely, this requires replacing $\rho_i$ with $\rho(x)$, where $x$ is a continuous analogue of $i$. Thus, we can rewrite Eq.~(\ref{eq:sink}) into 
\beq
\frac{\partial J(x)}{\partial x}=-\frac{k_d\sigma(x)}{s},
\eeq
where $s$ is the site size. For tractability, we assume a multiplicative expression for current. This way, we rewrite equation above into
\beq
\frac{\partial \nu(x)}{\partial x}\left[	1-\rho(x)	\right]-\nu(x)\frac{\partial \rho(x)}{\partial x}=-\frac{\left[\rho(x)-\nu(x)\right]k_d}{sp}\label{eq:diff_cont}.
\eeq
Furthermore, the steady-state form of Eq.~(\ref{eq:sigmaWsink}) in continuous limit reads
\begin{eqnarray}
0&=&k_-\nu(x) +\left[\nu(x) - s \frac{\partial \nu(x)}{\partial x}\right]\left[1-\rho(x)\right]p-\label{eq:sigmaWsinkContinuous}\\
&-&\left[\rho(x)	-	\nu (x)\right] \left\{k_+\left[1-\varphi \frac{1}{L}\int_0^L \nu(x){\rm d}x\right] +k_d\right\}.\nonumber
\end{eqnarray}
In this equation, we replaced the average density of token-bound particles $\langle \nu\rangle$ with its continuous analogue, {\sl i.e.,} $\langle \nu\rangle=(1/L)\int_0^L \nu(x){\rm d}x$. Equations (\ref{eq:diff_cont}) and~(\ref{eq:sigmaWsinkContinuous}) constitute a system of integro-differential equations that describe the density profile and the current. General solution of this system is beyond the scope of this work: below we analyze just $\lim_{k_d\to 0}$ to demonstrate how continuous limit corresponds to the results from before.

In this limit, Eq.~(\ref{eq:diff_cont}) becomes:
\beq
\frac{\partial \nu(x)}{\partial x}\left[	1-\rho(x)	\right]=\nu(x)\frac{\partial \rho(x)}{\partial x},
\eeq
which we rewrite into:
\beq
\frac{\partial\nu(x)}{\partial x} \frac{1}{\nu(x)}=\frac{\partial \rho(x)}{\partial x} \frac{1}{1-\rho(x)}.
\eeq
Using an ansatz $\nu(x)=\zeta \rho(x)$, we arrive to $\rho=1/2$, which is density obtained in maximal-current phase of the standard TASEP with open boundaries.

\bibliography{TASEP_stochastic.bib}

\end{document}